\documentclass[twocolumn]{aastex631} 

\usepackage[normalem]{ulem}

\newcommand\aastex{AAS\TeX}

\newcommand{\oiii}{[OIII]$_{88\mu m}\,$}

\newcommand{\ojwst}{[OIII]$\lambda$5007\AA}
\newcommand{\oojwst}{[OIII]$\lambda\lambda$4959,5007\AA}

\newcommand{\Hb}{H$\mathrm{\beta}$}
\newcommand{\ciii}{CIII]$\lambda\lambda$1907,9\AA}

\begin{document}

\title{Detection of \oiii in JADES-GS-z14-0 at z=14.1793} 

\author[0000-0001-9746-0924]{Sander Schouws}
\affil{Leiden Observatory, Leiden University, NL-2300 RA Leiden, Netherlands}

\author[0000-0002-4989-2471]{Rychard J. Bouwens}
\affil{Leiden Observatory, Leiden University, NL-2300 RA Leiden, Netherlands}

\author[0000-0003-2000-3420]{Katherine Ormerod}
\affil{Astrophysics Research Institute, Liverpool John Moores University, 146 Brownlow Hill, Liverpool L3 5RF, United Kingdom}

\author[0000-0001-7768-5309]{Renske Smit}
\affil{Astrophysics Research Institute, Liverpool John Moores University, 146 Brownlow Hill, Liverpool L3 5RF, United Kingdom}

\author[0000-0002-4205-9567]{Hiddo Algera}
\affil{Hiroshima Astrophysical Science
Center, Hiroshima University, 1-3-1 Kagamiyama, Higashi-Hiroshima, Hiroshima 739-8526, Japan}
\affil{National Astronomical Observatory of Japan, 2-21-1, Osawa, Mitaka, Tokyo, Japan}

\author[0000-0002-2906-2200]{Laura Sommovigo}
\affil{Center for Computational Astrophysics, Flatiron Institute, 162 Fifth Avenue, New York, NY 10010, USA}

\author[0000-0001-6586-8845]{Jacqueline Hodge}
\affil{Leiden Observatory, Leiden University, NL-2300 RA Leiden, Netherlands}

\author[0000-0002-9400-7312]{Andrea Ferrara}
\affil{Scuola Normale Superiore, Piazza dei Cavalieri 7, 50126 Pisa, Italy}

\author[0000-0001-5851-6649]{Pascal A. Oesch}
\affil{Departement d’Astronomie, Universit\'e de Gen\'eeve, 51 Ch. des Maillettes, CH-1290 Versoix, Switzerland}
\affil{Cosmic Dawn Center (DAWN), Copenhagen, Denmark}
\affil{Niels Bohr Institute, University of Copenhagen, Jagtvej 128, DK-2200 Copenhagen N, Denmark}

\author[0009-0009-2671-4160]{Lucie E. Rowland}
\affil{Leiden Observatory, Leiden University, NL-2300 RA Leiden, Netherlands}

\author[0009-0005-6803-6805]{Ivana van Leeuwen}
\affil{Leiden Observatory, Leiden University, NL-2300 RA Leiden, Netherlands}

\author[0000-0001-7768-5309]{Mauro Stefanon}
\affil{Leiden Observatory, Leiden University, NL-2300 RA Leiden, Netherlands}

\author[0000-0003-2164-7949]{Thomas Herard-Demanche} 
\affil{Leiden Observatory, Leiden University, NL-2300 RA Leiden, Netherlands}

\author[0000-0001-7440-8832]{Yoshinobu Fudamoto} 
\affiliation{Center for Frontier Science, Chiba University, 1-33 Yayoi-cho, Inage-ku, Chiba 263-8522, Japan}

\author[0000-0001-8887-2257]{Huub R\"ottgering}
\affil{Leiden Observatory, Leiden University, NL-2300 RA Leiden, Netherlands}

\author[0000-0002-4389-832X]{Paul van der Werf}
\affil{Leiden Observatory, Leiden University, NL-2300 RA Leiden, Netherlands}

\begin{abstract}

We report the first successful ALMA follow-up observations of a secure $z > 10$ JWST-selected galaxy, by robustly detecting ($6.6\sigma$) the [OIII]$_{88\mu m}\,$ line in JADES-GS-z14-0 (hereafter GS-z14). The ALMA detection yields a spectroscopic redshift of $z=14.1793\pm0.0007$, and increases the precision on the prior redshift measurement of $z=14.32_{-0.20}^{+0.08}$ from NIRSpec by $\gtrsim$180$\times$. Moreover, the redshift is consistent with that previously determined from a tentative detection ($3.6\sigma$) of CIII]$_{1907,1909}$ ($z=14.178\pm0.013$), solidifying the redshift determination via multiple line detections. We measure a line luminosity of $L_\mathrm{[OIII]88} = (2.1 \pm 0.5)\times10^8\,L_\odot$, placing GS-z14 at the lower end, but within the scatter of, the local $L_\mathrm{[OIII]88}$-star formation rate relation. No dust continuum from GS-z14 is detected, suggesting an upper limit on the dust-to-stellar mass ratio of $< 2 \times 10^{-3}$, consistent with dust production from supernovae with a yield $y_d < 0.3\,M_\odot$. Combining a previous JWST/MIRI photometric measurement of the \oojwst~ and H$\beta$ lines with {\sc Cloudy} models, we find GS-z14 to be surprisingly metal-enriched ($Z\sim0.05 - 0.2\,Z_\odot$) a mere $300\,\mathrm{Myr}$ after the Big Bang. The detection of a bright oxygen line in GS-z14 thus reinforces the notion that galaxies in the early Universe undergo rapid evolution.

\end{abstract}


\section{Introduction} \label{sec:intro}
The discovery and spectroscopic confirmation of galaxies at $z>10$ has recently become possible due to the groundbreaking capabilities offered by the James Webb Space Telescope (\textit{JWST}) \citep[e.g.][]{Curtis-Lake_2022, Robertson_2023, Bunker_2023, carniani2024, Finkelstein_2023, zavala2024}. In particular JADES-GS-z14-0 (hereafter GS-z14) was recently spectroscopically confirmed to be the most distant known galaxy at $z_\mathrm{spec}= 14.32_{-0.20}^{+0.08}$, less than 300 Myr after the Big Bang \citep{carniani2024}. Notably GS-z14 is also very luminous with $M_{UV}=-20.81\pm0.16$, which makes it the second most luminous $z>8$ galaxy with a spectroscopic redshift; only GN-z11 \citep{Oesch_2016,bunker2023} is more luminous by a factor $\sim$2$\times$.

Moreover, in contrast to other $z>10$ galaxies, the rest-frame UV morphology of GS-z14 is extended and not highly concentrated. This implies that the luminosity is dominated by a spatially extended stellar population as opposed to an active galactic nucleus (AGN). The existence of objects like GS-z14 suggests a much more rapid build-up of galaxies in the very early universe than previously expected \citep{carniani2024}. GS-z14 enables a unique opportunity to study this rapid build-up in detail \citep{Ferrara2024a}.

The high luminosity of GS-z14 makes it an exceptional target for multi-wavelength follow-up observations capable of revealing its physical conditions. Here we present new observations of GS-z14 targeting the luminous \oiii far-infrared line with the the Atacama Large Millimeter/submillimeter Array (ALMA).

The [OIII] 88-micron fine-structure line is one of the dominant coolants of the ISM. It originates from ionized gas in HII regions \citep{Cormier2015}, where ionizing radiation from young, massive stars has stripped electrons from oxygen atoms. 
Both semi-analytical models \citep{Yang_2020,vallini2021,vallini2024} 
and radiative transfer post-processed large box \citep{Moriwaki_2018}, cosmological zoom-in \citep{katz2017,arata2020,katz2022,pallottini2022,kohandel2023,nakazato2023} 
hydrodynamical simulations have been employed to predict and interpret \oiii emission in high redshift sources. The consensus is that bright \oiii emission is generally associated with hard ionisation fields \citep{Yang_2020,pallottini2022}, high ionization parameters \citep{Moriwaki_2018, arata2020, vallini2024, nakazato2023}, and low metallicities \citep{Katz_2019,vallini2024}. 
These physical properties are likely to be ubiquitous in very actively star-forming galaxies at high redshifts, making this line the ideal target for ALMA follow-up observations of JWST-selected $z>10$ galaxies.  

In the last two years, ALMA follow-up observations have been performed for several high-redshift galaxy candidates. GHZ2 \citep{castellano2022,naidu2022,Donnan_2022,harikane2022,bouwens2023}, with a photometric redshift of $z = 11.96-12.42$, was targeted with an ALMA Band 6 search for \oiii by \citet{bakx2023}, who determined a $5\sigma$ upper limit of $\log (L_{\rm [O III]} / L_\odot) < 1.7 \times 10^8$ (see also \citealt{Popping_2023}). GHZ2 was later spectroscopically confirmed at $z=12.34$ \citep{castellano2024}, with \citet{zavala2024} estimating an SFR of $9\pm3\,M_\odot\,\mathrm{yr}^{-1}$ from the H$\alpha$ line. Following its spectroscopic confirmation, \citet{zavala2024} re-examined the ALMA observations of GHZ2 presented in \citet{bakx2023}, but were not able to identify a plausible $>$5$\sigma$ \oiii detection at the expected frequency. 


An ALMA line scan was also performed on GHZ1, with a photometric redshift of $z\approx10.6$ \citep{Treu_2023}, SFR of $36.3^{+54.5}_{-26.8} \ M_{\odot} \ \mathrm{yr}^{-1}$ and stellar mass $\log (M_\star / M_\odot) = 9.1^{+0.3}_{-0.4}$ \citep{Santini_2023}, yielding a marginal \oiii signal near its JWST position \citep{Yoon_2023} and an upper limit of $L_{\rm [OIII]} < 2.2 \times 10^8 \ L_\odot$ \textcolor{black}{(5$\sigma$)}. Another high-redshift candidate, HD1 at $z = 13.3$ \citep{Harikane_2023} was observed in ALMA Bands 4 and 6 targeting [CII]$_{158\mu m}$ and [OIII]$_{88\mu m}$, respectively, but the lines were not detected \citep{kaasinen2023}. HD1 was later shown to be a low-redshift interloper at $z=4.0$ \citep{harikane2024}. Finally, S5-z17-1, identified in JWST ERO data, showed a potential 5.1$\sigma$ detection at 338.726 GHz, possibly corresponding to [OIII]$_{52\mu m}$ at $z = 16$ \citep{Fujimoto_2022}, suggesting an SFR $<$ 120 $M_\odot \ \mathrm{yr}^{-1}$. However, the high-redshift nature of this galaxy has not been conclusively established.

Overall, these upper limits and non-detections indicate possible lower redshift solutions or insufficient sensitivity in the requested observations \citep{bakx2023,Furlanetto_2023,kaasinen2023}. However, the burstiness of SF in such high-redshift galaxies, as well as the impact of feedback processes on galaxy spectra and visibility, have also been suggested as physical motivations for the lack of ALMA detections at $z>10$ \citep{nakazato2023,kohandel2023}. 

Throughout this paper we assume a standard $\mathrm{\Lambda}$CDM cosmology with $H_0=70$ km s$^{-1}$ Mpc$^{-1}$, $\Omega_m=0.3$ and $\Omega_{\Lambda}=0.7$. Magnitudes are presented in the AB system \citep{oke_gunn_1983ApJ...266..713O}. For star formation rates and stellar masses we adopt a Chabrier IMF \citep{chabrier2003}. Error-bars indicate the $68\%$ confidence interval unless specified otherwise. All measured and derived physical quantities are corrected for gravitational lensing by a factor of 1.17$\times$ \citep{carniani2024}. Logarithms use base 10 unless specified otherwise.  

\begin{figure*}[th!]
\epsscale{1.15}
\plotone{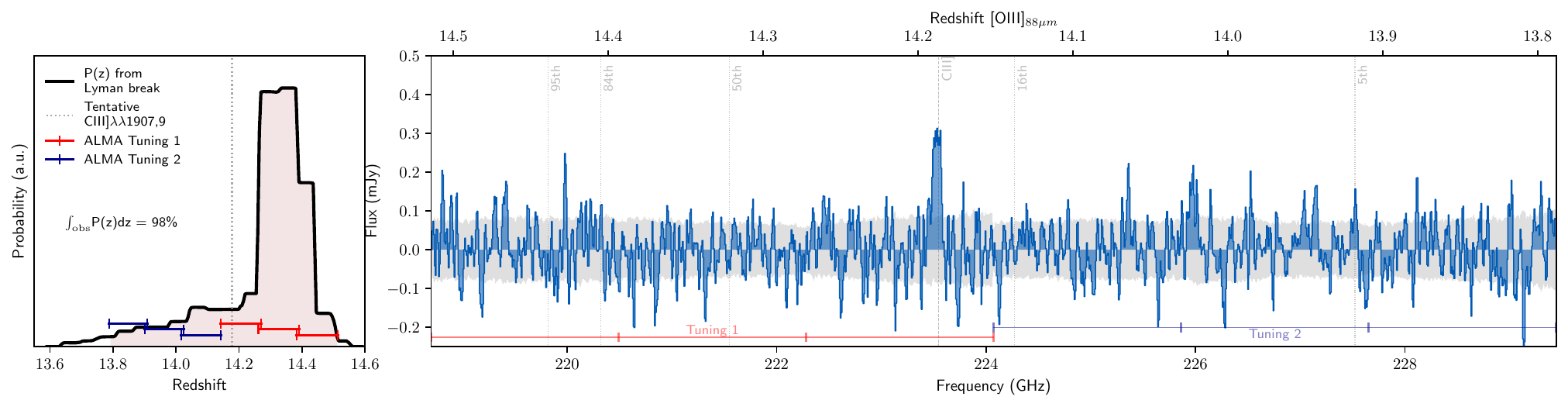} 
\caption{\textit{Left Panel:} The ALMA observations cover 98\% of the redshift probability distribution derived from the Lyman break as observed by NIRSpec PRISM. We use two tunings with three spectral windows each, tuning 1 and 2 cover 83\% and 15\% of the probability distribution respectively. \textit{Right Panel:} Flux constraints extracted at the position of GS-z14 (\textit{blue histogram}) vs. frequency from the 2023.A.00037.S spectral scan observations.  The grey contours indicate the $1\sigma$ uncertainties.  For context we show the expected redshift from the tentative CIII]$_{1907,1909}$ line from NIRSpec as well as the 5, 16, 50, 86 and 95th percentiles of the P(z) distribution.  There is only one feature in the line scan that can be confidently identified as \oiii line emission and it occurs at a frequency of 223.528$\pm$0.009 GHz. } \label{fig:scan} 
\end{figure*}

\section{Observations and Data-Reduction} \label{sec:observations}

\subsection{JWST} 
GS-z14 was discovered in deep imaging of the GOODS-South field obtained by the JWST Advanced Deep Extragalactic Survey \citep[JADES;][]{Eisenstein_2023} and the First Reionization Epoch Spectroscopic COmplete Survey \citep[FRESCO;][]{Oesch_2023}. GS-z14 was initially flagged as a likely low redshift interloper \citep{Hainline_2023, Williams_2023} due to a nearby foreground galaxy at $z=3.475$ with a separation of only 0.4” and due to its high luminosity. However, further analysis including additional medium-band observations favored a high redshift solution \citep{Robertson_2023}.

Deep follow-up spectroscopy with NIRSpec presented in \citet{carniani2024} shows a strong break at $\sim$1.85$\mu$m, consistent with a Lyman break at $z\sim14$. The profile of the Lyman break is sensitive to absorption of hydrogen along the line of sight, neutral gas in the galaxy and environment, velocity offsets, the presence of Lyman alpha emission and possible ionized bubbles. Accounting for these effects, \citet{carniani2024} determined a spectroscopic redshift of z = 14.32$^{+0.08}_{-0.20}$. 

In contrast to other $z>10$ sources, the spectrum of GS-z14 does not contain strong rest-frame UV emission lines. Only \ciii\, is tentatively detected with a significance of 3.6$\sigma$ at $z=14.178\pm0.013$, consistent with the redshift obtained from the Lyman Break. \citet{carniani2024} discuss a number of mechanisms that could be responsible for the lack of strong emission lines in the rest-frame UV, ranging from a sudden quenching of the star formation, very low or high metallicities (Z$<$0.05Z$_{\odot}$ or Z$\sim$Z$_{\odot}$), a high escape fraction of ionizing photons and a difference in the dominant ionizing flux.

Interestingly, GS-z14 is also detected by MIRI in the F770W filter \citep{helton2024}. This filter covers the rest-frame optical emission of GS-z14, corresponding to wavelengths of 4.4 to 5.7$\mu$m. This part of the spectrum contains the strong \oojwst\, and \Hb\, emission lines, and indeed the observed flux is significantly stronger than observed in the adjacent F444W filter, implying a significant contribution from the emission lines (constituting about $1/3$ of the total F770W flux).

SED fitting of GS-z14 performed by \citet{helton2024} shows that the photometry is consistent with a stellar mass of log(M$_{*}$/M$_{\odot}$)=$8.7_{-0.4}^{+0.5}$. The models estimate that most of this stellar mass was formed relatively recently, with a mass-weighted age of $\sim$20 Myr. \citet{helton2024} determine the current ($<10\,\mathrm{Myr}$) star formation rate of GS-z14 to be 25$^{+6}_{-5}$ M$_{\odot}$ yr$^{-1}$---consistent with the measurement of $22\pm6\,M_\odot$ yr$^{-1}$ from \citet{carniani2024}. Combined with the measured size of 260$\pm$20 pc, this implies a high star formation rate surface density of $\sim$64 M$_{\odot}$ yr$^{-1}$ pc$^{-2}$, comparable to intense starbursting galaxies in the local universe \citep{Genzel_2010}. The UV slope $\beta_{UV} =- 2.20 \pm 0.07$ indicates the presence of a moderate amount of dust in addition to the very young stellar population, with a visual extinction of A$_{V}$=0.31$^{+0.14}_{-0.07}$ \citep{carniani2024}. Finally, the metallicity is poorly constrained owing to a lack of detected emission lines but is expected to be low (Z = 0.014$^{+0.052}_{-0.012}$ Z$_{\odot}$; \citealt{carniani2024}).

\vspace{0.75cm}
\subsection{ALMA} 
\vspace{-0.1cm}
The ALMA Band 6 follow-up observations targeting the \oiii line were obtained as part of a Cycle 10 Director's Discretionary Time (DDT) program (\#2023.A.00037.S, PI: Schouws). An observing set-up using two tunings with three spectral windows each was utilized to maximize the coverage of the redshift likelihood distribution derived from the spectroscopic Lyman break. This results in a continuous frequency coverage ranging from 218.70 GHZ to 229.45 GHz, corresponding to a redshift range of $z=13.79$ to 14.51. We show in Figure \ref{fig:scan} that the observing strategy covers 98\% of the P(z).

The observations for Tuning 2 were carried out between 15 and 16 August 2024 in good weather conditions (PWV=0.67mm), achieving the requested sensitivity in 2.8 hours.  Tuning 1 was observed between 7 and 8 September 2024 for 2.8 hours in excellent conditions (PWV=0.30mm). Both tunings were observed with different array configurations due to scheduling constraints.  Tuning 1 was observed in C-4 and Tuning 2 in C-5 with baselines of 15$-$500 and 15$-$919 m,  \textcolor{black}{resulting in synthesized beams of $1.09''\times 0.81''$ and $0.57''\times 0.49''$ respectively when using natural weighting.} 


The ALMA data were reduced and calibrated following the standard ALMA pipeline procedures with the Common Astronomy Software Applications (v6.5.4-9) \citep[\textsc{Casa;}][]{Hunter_2023}. \textcolor{black}{The amplitude and phase calibrations were performed using J0334-4008 and J0348-2749, respectively. Both calibrators are part of the ALMA Calibrator Source Catalogue and are therefore considered reliable for Band 6 observations. Under typical conditions, the absolute flux calibration accuracy for Band 6 is 5–10\% \citep{ALMA_Tech_Handbook_C10}. The data were well behaved, with no bad antennas or channels requiring flagging.}

The resulting calibrated measurement sets were time-averaged in bins of 30 seconds to reduce the data-size, after carefully verifying that time-average smearing does not impact our results \citep[e.g.][]{thompson2017}.

Imaging of the calibrated visibilities was performed with natural weighting using the \textsc{tclean} task in \textsc{Casa}, cleaning to a depth of 2$\sigma$ using automasking \citep{kepley2020}. We use a pixel scale of 0.076$''$ to properly sample the synthesized beam, which has a FWHM of $1.08''\times 0.80''$ at the frequency of the \oiii line.


\begin{figure*}[th!]
\epsscale{1.15}
\plotone{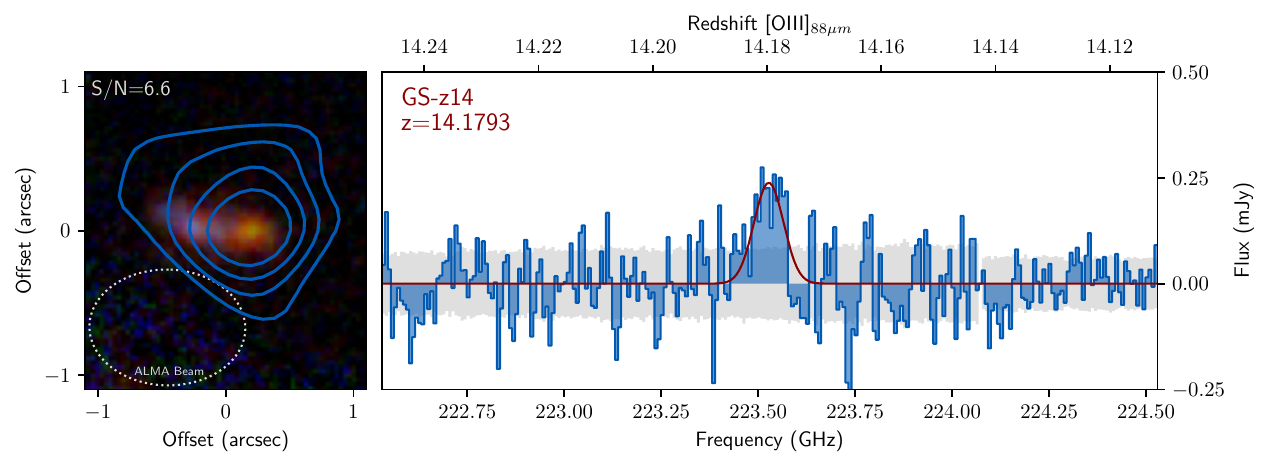} 
\caption{Detection of \oiii in GS-z14 at $z=14.1793\pm0.0007$. \textit{Left Panel:} Contours showing the \oiii emission (2, 3, 4 and 5$\sigma$) overlaid on \textcolor{black}{an RGB image based on F150W, F200W and F444W} imaging \citep[][]{Eisenstein_2023,Eisenstein_2023_JOF}. \textcolor{black}{The object with a bluer colour to the left of GS-z14 is a z=3.475 foreground galaxy \citep[][]{carniani2020}.} The \oiii emission is detected at a peak significance of 6.6$\sigma$ in the collapsed data-cube \textcolor{black}{and well centered on GS-z14}. \textit{Right Panel:} The spectrum of the \oiii$\,$line (\textit{blue bars}) extracted from the $>$3$\sigma$ emission region in the moment-0 map. The red line shows the Gaussian fit used to measure the spectroscopic redshift and FWHM of the line.  The grey shaded region indicates the $1\sigma$ uncertainties.} \label{fig:detail}
\end{figure*}

\vspace{-0.1cm}
\subsection{Line Search} 
\vspace{-0.1cm}

We perform a blind search for emission lines using an algorithm similar to the one used by \citet{bethermin2020} for the ALPINE survey \textcolor{black}{\citep{lefevre2020,faisst2020}} and Schouws et al.\ in prep. for the REBELS survey \textcolor{black}{\citep{bouwens2022}}. The algorithm loops over all channels and collapses moment maps over a range of 75 to 350 km s$^{-1}$ in steps of 1 channel ($\sim$10 km s$^{-1}$). For each moment map, we identify the significant ($>$3$\sigma$) peaks and add the results to a large list of features. This list is then pruned by removing duplicates within 2$\times$FWHM and 1.5$\times$ the beamsize. We also perform the search on the negative moment map, which is useful to characterize the noise properties. In Figure \ref{fig:scan}, we show a blind extraction of the ALMA spectrum at the location of GS-z14 using a 0.5" aperture.

The most significant line candidate we extract within a 0.5" radius of the JWST position of GS-z14 is detected with a peak significance of 6.6$\sigma$ and separation of 0.12".  The candidate line is consistent with being spatially coincident with GS-z14 in the JWST/NIRCam observations given expected 0.18" positional uncertainties following from a 6.6$\sigma$ line detection and 1.08$"$ beam.\footnote{https://help.almascience.org/kb/articles/what-is-the-absolute-astrometric-accuracy-of-alma}


\begin{figure}[b!]
\vspace{2.0cm}
\epsscale{1.0}
\plotone{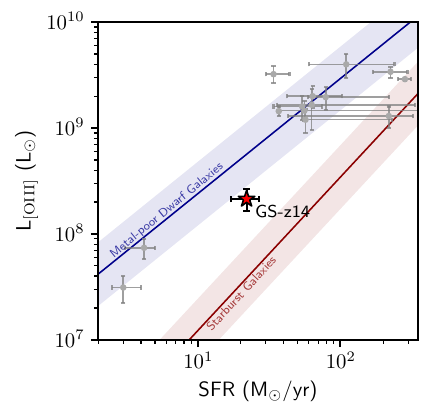} 
\caption{The luminosity of \oiii in GS-z14 is consistent with the local relation for metal-poor dwarf galaxies found by \citet{delooze2014}. For context we also show the local relation for starburst galaxies \citep{delooze2014} as well as a compilation of $z>6.5$ galaxies \citep{bakx2020,carniani2020,harikane2020,akins2022,Witstok_2022,algera2024,fujimoto2024a}. Even as early as $z=14.1793\pm0.0007$ and for high redshift galaxies in general, the relation for metal-poor dwarf galaxies seems to be a good fit (at $\mathrm{SFR}\lesssim100\,M_\odot\,\mathrm{yr}^{-1}$). } \label{fig:OIII-sfr} 
\end{figure}

\vspace{-0.25cm}
\section{Results} \label{sec:results}
\vspace{-0.1cm}

\begin{figure*}[t!]
\vspace{-0.5cm}
\epsscale{1.15}
\plotone{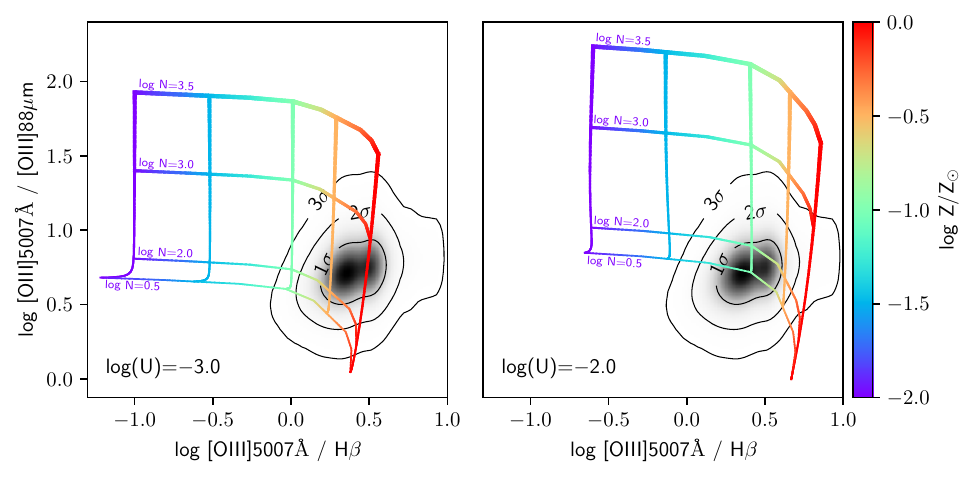} 
\caption{The ALMA and MIRI observations constrain the emission line ratios of GS-z14 to the dark shaded region in the figure highlighted by the 1 and 2$\sigma$ contours. ISM modeling as shown by the coloured grid highlights the dependence on metallicity and density. This indicates that GS-z14 is consistent with a relatively high metallicity (Z$>$0.1 Z$_{\odot}$) and moderate to low density log(N)$<3.0$cm$^{-3}$.  There is a further minor dependence on the ionization parameter that shifts and deforms the grid, but does not impact the conclusions. For context, SED fitting of GS-z14 finds $\log (U)=-2.5\pm0.5$ \citep{helton2024}.} \label{fig:ISM}
\end{figure*}

\subsection{\oiii in GS-z14} 
We have identified an emission line with a signal-to-noise of $6.6\sigma$ within $0.12''$ of GS-z14 at 223.528$\pm$0.009 GHz. This corresponds to \oiii at $z=14.1793\pm0.0007$, consistent with both the redshift derived from the Lyman break observed by JWST and also the tentative 3.6$\sigma$ detection of CIII]$_{1907,1909}$ at 2.89$\mu$m \citep{carniani2024}.  In particular, the consistency of our new redshift determination from \oiii$\,$ with the earlier redshift estimate $z=14.178\pm0.003$ from the tentative CIII]$_{1907,1909}$ doublet \citep{carniani2024} greatly increases our confidence in the robustness of the current redshift determination given the availability of multiple line detections.  We show the contours of the \oiii emission overlaid on \textcolor{black}{an RGB image based on F150W, F200W and F400W} imaging and a S/N-optimized extraction of the spectrum in Figure \ref{fig:detail}. 

We measure the integrated line flux using the moment-zero map of the emission line, including all channels that fall within 2$\times$ the FWHM of the line. The FWHM is determined in an iterative process; starting from an estimate of the FWHM we collapse a moment-zero map and extract a 1d spectrum, which is extracted by including all pixels with S/N$>$3$\sigma$ on the moment-zero map. A new FWHM is then measured using this 1d spectrum by fitting a Gaussian. This new FWHM is then used to collapse a new moment-zero map for the next iteration. A stable FWHM is achieved in less than 10 steps \citep{schouws2022}.

The final FWHM we measure is 136$\pm$31 km s$^{-1}$. Assuming that GS-z14 is dispersion-dominated and r$_\mathrm{[OIII]}$ = r$_e$ = 260 pc, this implies a dynamical mass of (1.0$\pm$0.5)$\times$10$^9$ M$_{\odot}(\mathrm{sin}\, i)^2$. At the current resolution we do not see evidence for a velocity gradient.

We measure an integrated flux of 39$\pm$10 mJy$\cdot$km s$^{-1}$ this corresponds to a \oiii luminosity of $(2.1\pm0.5)\times10^8\,$L$_{\odot}$ \citep{Solomon_1992} (after correction for lensing magnification; \citealt{carniani2024}). This places GS-z14 a factor of $\sim2\times$ below the local relation between  L$_\mathrm{[OIII]}$ and SFR for metal-poor dwarf galaxies from \citet{delooze2014}, albeit nearly within the scatter (Figure \ref{fig:OIII-sfr}). The general consistency with the $z=0$ relation for a galaxy a mere $\sim300\,$Myr after the Big Bang suggests that GS-z14 has undergone rapid evolution, as discussed in detail in Section \ref{sec:ism_constraints}.


\vspace{-0.35cm}
\subsection{Dust Continuum} 

We do not detect the dust continuum from GS-z14, with a formal 90$\mu$m continuum limit of $< 15.1\,\mu\mathrm{Jy\,beam}^{-1}$ (3$\sigma$).  We provide an upper limit on its dust mass \textcolor{black}{assuming an optically thin modified black body with a dust temperature of $T_\mathrm{d} = 60\,\mathrm{K}$ and $\beta_\mathrm{IR} = 2.03$ for the dust SED. The assumed} temperature is consistent with extrapolations from theoretical models (e.g., \citealt{liang2019,sommovigo2022}) and observational trends (e.g., \citealt{schreiber2018,faisst2020_tdust,sommovigo2021,sommovigo2022,sommovigo2022b,Witstok23_dust}) to $z\approx14$. \textcolor{black}{We note that the precise evolution of dust temperature is still uncertain and debated \citep[e.g.][]{sommovigo2022b}. However, our assumption of a relatively high dust temperature is also supported by the inverse correlation between metallically and dust temperature noted by \citet{sommovigo2022b} and increasing temperature of the CMB background at high redshift\footnote{For context, the CMB background at $z=14.2$ has a temperature of $T_{\mathrm{CMB}}=41.4$K.}.}

We follow \citet{Ferrara2024b} by adopting the \citet{weingartner2001} dust model with $\kappa_{88} = 34.15\,\mathrm{cm}^{-2}\,\mathrm{g}^{-1}$ at rest-frame $88\,\mu\mathrm{m}$. This yields a (lensing-corrected) upper limit on the dust mass and infrared luminosity (\textcolor{black}{integrated across $8-1000\,\mu\mathrm{m}$}) of $\log(M_\mathrm{d} / M_\odot) < 6.0$ and $\log(L_\mathrm{IR} / L_\odot) < 11.1$, respectively. The latter corresponds to an obscured SFR of $\mathrm{SFR}_\mathrm{IR} < 14\,M_\odot\,\mathrm{yr}^{-1}$ which, combined with the unobscured SFR of GS-z14 inferred by \citet{carniani2024}, implies an obscured fraction of $f_\mathrm{obs} < 0.66$. Finally, using the stellar mass of GS-z14 inferred by \citet{helton2024}, we infer a dust-to-stellar mass ratio  $\xi_d \equiv M_\mathrm{d} / M_\star < 1.9 \times 10^{-3}$.

\textcolor{black}{The above estimates require the assumption of a dust temperature. Assuming a lower temperature of $T_\mathrm{d} = 45\,\mathrm{K}$, which would imply no significant evolution of the dust temperature from $z\sim4$ \citep[e.g.][]{faisst2020_tdust}, results in a dust mass of  $\log(M_\mathrm{d} / M_\odot) < 6.8$. Meanwhile, a significantly higher temperature temperature of $T_\mathrm{d} = 100\,\mathrm{K}$ would result in a dust mass of  $\log(M_\mathrm{d} / M_\odot) < 5.3$.} In Section \ref{sec:dust_formation}, we discuss this in further detail, focusing on the Attenuation Free Model by \citet{Ferrara2024b} which self-consistently predicts a dust temperature for GS-z14 based on the observed V-band attenuation and the spatial extent of the dust.



\section{Discussion} \label{sec:discussion}

\subsection{Constraints on the ISM of GS-z14} \label{sec:ism_constraints}
Luminosity ratios of emission lines are invaluable diagnostic tools to probe the conditions of the ISM in galaxies. Although the NIRSpec spectroscopy of GS-z14 does not show significant emission lines, its detection in the MIRI F770W band places a constraint on the combined \oojwst+\Hb\, flux. Following \citet{helton2024} we assume that the flux excess in F770W is 27.5$\pm$5.6 nJy based on a flat underlying continuum emission. 

The \ojwst\, luminosity (hereafter [OIII]) can subsequently be calculated based on an assumed [OIII]/\Hb\, ratio and be compared to the luminosity of OIII as measured by ALMA. We repeat this calculation 10$^6\times$, drawing random [OIII]/\Hb\, ratios from probability distribution from the SED fitting by \citet{helton2024}, where we combine the probability distributions from the three star formation history assumptions. \textcolor{black}{This distribution has a median ratio of 2.5$^{+0.9}_{-0.6}$. As noted by \citet{helton2024}, this is lower than the median [OIII]/\Hb\, at $z\sim8.0$ which is 6.0$^{+2.4}_{-2.9}$ in JADES \citep{helton2024}, 6.4$\pm$0.9 in FRESCO \citep{meyer2024} and 7.2$^{+2.6}_{-2.4}$ in PRIMAL \citep{Heintz2025}.}

In this process we also account for the uncertainty in the line luminosities from ALMA and JWST by drawing random values from a Gaussian error distribution.

The resulting constraints on the [OIII]/\oiii\, versus [OIII]/\Hb\, line ratios are shown in Figure~\ref{fig:ISM}. For context we show a grid of ISM models for a large range of conditions based on {\sc 
 Cloudy} models \citep{ferland2017}. The models consist of an HII region that smoothly transitions to a Photo Dissociation Region (PDR) until a fixed optical depth (A$_V$=10) in a plane parallel geometry. For more details on the model we refer to \citet{Witstok_2022}. We note that the ionization parameter of GS-z14 has been loosely constrained by \citet{helton2024} to be $\log(U_\mathrm{ion}) = -2.5 \pm 0.5$, and we therefore show two sets of {\sc Cloudy} models bracketing this range, at a fixed $\log(U_\mathrm{ion}) \in (-3.0, -2.0)$. We focus on the grid with an ionization parameter $\log(U_\mathrm{ion}) = -2$, as \citet{kohandel2023} predict the most \oiii-luminous galaxies at $z \gtrsim 10$ have high ionization parameters. \textcolor{black}{Moreover, if the assumed [OIII]/\Hb\, is underestimated, the observed line ratios can only be reproduced by higher values of the ionization parameter.} 

While the uncertainties are substantial, given the existence of only a photometric detection of the \oojwst~ line, the \citet{Witstok_2022} models suggest a relatively high metallicity of $Z \sim 0.1\,Z_\odot$, in combination with a moderate-to-low electron density ($\log N \lesssim 10^{2.5}$). This is on the low end of, albeit still consistent with, the distribution of electron densities found for $z\gtrsim5$ galaxies, which typically show values of $\log(N) \sim 2 - 3$ (e.g., \citealt{isobe2023}). However, for more accurate constraints on the ISM conditions of GS-z14, direct spectroscopic detections of additional oxygen lines are crucial, for example through MIRI spectroscopy.

\begin{figure}[h]
\plotone{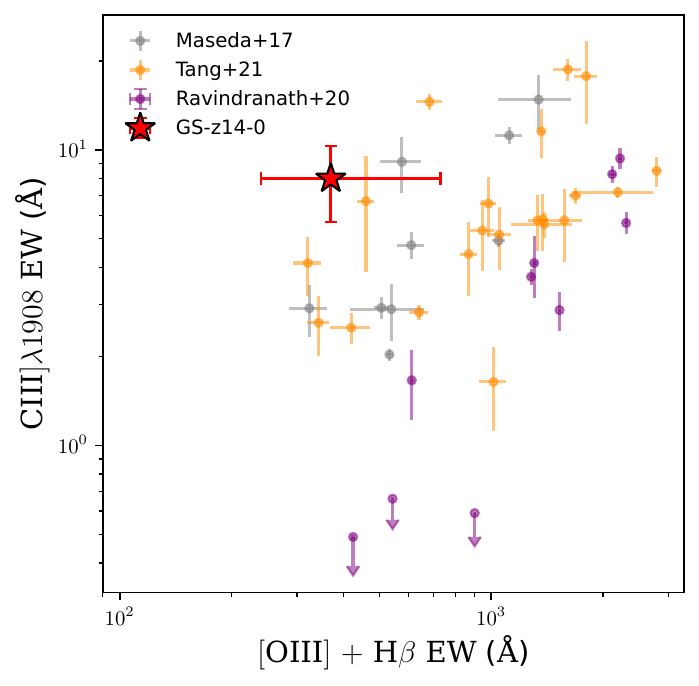} 
\caption{The rest-frame equivalent width of \ciii\ as a function of EW$_0$([OIII]+\Hb) \citep{carniani2024,helton2024}. GS-z14-0 is broadly consistent with measurements from the literature at $z=1-3$ \citep[grey, purple and orange points;][]{Maseda2017,Ravindranath_2020,Tang2021}.   } \label{fig:OIII_CIII_EW} 
\end{figure}

Intriguingly, the detection of \oiii at $z=14.1793\pm0.0007$ is consistent within 1$\sigma$ with the tentative detection of \ciii$\,$ reported by \citet{carniani2024} at $z=14.178\pm0.013$. \citet{carniani2024} derive a \ciii$\,$ rest-frame equivalent width (EW$_0$) of 8.0$\pm$2.3 \AA, which is relatively high compared to the EW$_0$([OIII]+\Hb) of 370$^{+360}_{-130}$ \AA$\,$ estimated by \citet{helton2024}, albeit consistent with the distribution found for galaxies at $z\sim0-4$ \citep[e.g.][]{Maseda2017,Ravindranath_2020,Tang2021}, see Figure \ref{fig:OIII_CIII_EW}. The measured equivalent widths are in agreement with predictions from {\sc Cloudy} at a metallicity of $Z\sim0.05-0.2\,Z_\odot$, and with an ionization parameter of $\log(U) \sim -2.5$, in agreement with \citet{helton2024}.

\begin{figure}[h]
\plotone{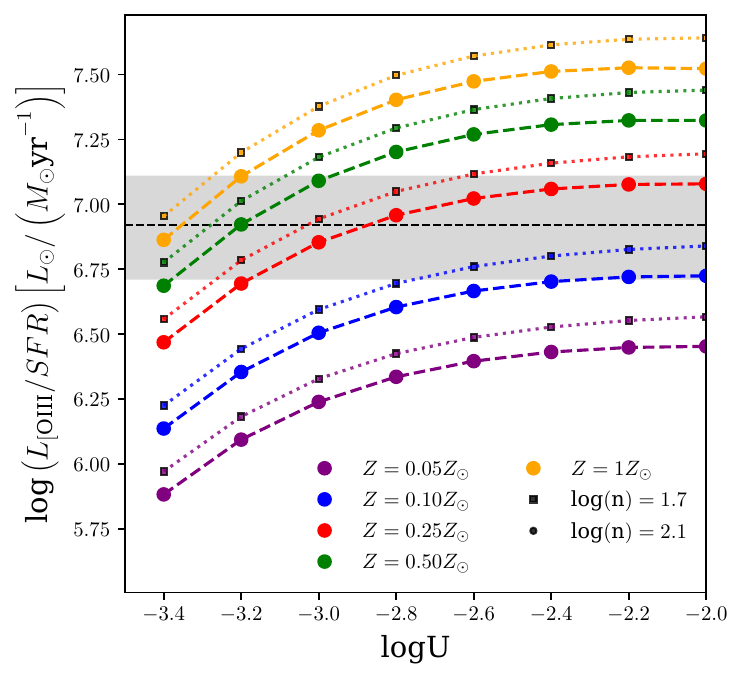} 
\caption{ {\sc Cloudy} modelling of the \oiii/SFR as a function of ionisation parameter for a range of metallicities and densities $\sim$50-100 cm$^{-3}$. Our model gives a lower limit on the metallicity of $\sim$10\% $Z_{\odot}$, with higher gas densities and older ages preferring higher metallicity estimates.   } \label{fig:loiii_sfr} 
\end{figure}

Moreover, \citet{JonesTucker2020} derive a calibration for the oxygen abundance using the \oiii\ line and star-formation rate. Using their calibration we derive 12 + $\log{\rm O^{++}/H^+}$=7.66$_{- 0.21}^{+0.19}$ ($\sim$6-14\% solar metallicity, using the solar abundance from \citet{asplund2009}). In Figure \ref{fig:loiii_sfr} we use a {\sc Cloudy} grid with a 1Myr old input stellar population to estimate the \oiii/SFR ratio as a function of ionisation parameter, for models with different metallicity and gas density. For modest gas density the measurements are consistent with 10\% solar oxygen abundance, however, higher gas densities or older ages for the input stellar population will give higher metallicity estimates (up to solar metallicity). 

The detection of both carbon and oxygen lines thus reinforce the notion that GS-z14 is already moderately chemically enriched. Adopting a fiducial $Z\sim0.05 - 0.2\,Z_\odot$ in combination with the stellar mass from \citet{helton2024}, GS-z14 falls onto the high-redshift mass-metallicity relation, which has been mapped out to $z\sim10$, and appears to show only mild evolution beyond $z\gtrsim 3$ (e.g., \citealt{curti2024}).

\subsection{Constraints on Dust Formation Processes}
\label{sec:dust_formation}

The build-up of dust in galaxies is a complex process involving multiple mechanisms of dust production and destruction, operating on different timescales. Supernova explosions (SNe) produce dust quickly, on timescales similar to the lifecycle of massive stars ($\sim$10 Myr), whilst Asymptotic Giant Branch (AGB) stars contribute to dust production on significantly longer timescales ($\sim$300 Myr, \citealt{schneider2024}). The contribution from grain growth in the ISM depends on the physical conditions, but is expected to be sub-dominant at extremely high redshifts and low gas-phase metallicities \citep{ferrara2016,Dayal_2022,Witstok23_dust,Markov_2024}.

At lower redshifts it is difficult to disentangle the impact of different contributions (e.g., \citealt{delooze2020,galliano2021}), but GS-z14 has formed sufficiently early in cosmic history that dust production through SNe is likely the dominant formation process. This makes GS-z14 an ideal test-bed for dust formation models in the early universe.

The efficiency of dust production from SNe can be inferred by measuring the dust-to-stellar mass ratio ($\xi_d$). This parameter is hard to interpret as it depends on both the expected number of SNe per unit of formed stellar mass ($\nu$), which in turn depends on the assumed Initial Mass Function (IMF), and on the net dust yield per SN event ($y_d$), $\xi_d = y_d \times \nu$. In local SN remnants $y_d$ values varying in the wide range from 0.01 to 1.1 M$_{\odot}$ have been directly measured \citep{Milisavljevic_2024}. Such values are broadly in agreement with most indirect high-z measurements based on dust-to-stellar mass ratios \citep{sommovigo2022,sommovigo2022b}. 

Utilizing the continuum non-detection of GS-z14, we can attempt to place an upper limit on $y_d$. Following \citet{michalowski2015} and adopting a Salpeter IMF, the limit of $\xi_d < 2 \times 10^{-3}$ suggests a yield of $y_d < 0.24\,M_\odot/$SN. 
This is consistent with a commonly adopted yield of $y_d\sim0.1\,M_\odot$/SN (e.g., \citealt{sommovigo2022,dayal2022}).
If we rely on such a value for $y_d\sim0.1\,M_\odot$/SN and assume that dust follows a similar spatial distribution as the stellar component, GS-z14 should be largely obscured (A$_{V}$$\sim$9.5, see also \citealt{Ferrara2024b}) and would not have been detected with JWST. However, the JWST data only reveal a relatively low visual extinction of A$_V$=0.31, implying a much (1 dex) lower dust mass of 5$\times$10$^4$ M$_{\odot}$ \textcolor{black}{with} $\xi_d$ $<$ 10$^{-4}$ \textcolor{black}{and $y_d < 0.015\,M_\odot/$SN} \citep{Ferrara2024b}. 

Clearly, SNe (and later on growth in the ISM) do produce tangible amounts of dust, as less than $500\,\mathrm{Myr}$ later, at $z\approx7$, dust appears widespread \citep{inami2022,Schouws_2022,Witstok23_dust,algera2023} -- both puzzlingly high dust-to-stellar mass ratios ($M_{\rm d}/M_{\star}\sim 0.01$; \citealt{algera2024}) and fully dust-obscured sources \citep{fudamoto2021} are observed at this epoch. In addition, features of the attenuation curve associated with the carbonaceous dust grains produced by SNe are observed at $z\sim 6.7$ \citep{Witstok2023_JWST}. 


The discrepancy between what is observed by JWST versus what is expected based on dust production from past SNe can be resolved by assuming that the majority of the dust has been removed from star-forming regions by radiation pressure-driven outflows. Such a scenario had been suggested in the Attenuation Free Model (AFM) presented in \citet{Ferrara2024b}. 

In their fiducial model, \citet{Ferrara2024b} suggest the dust in GS-z14 to have a typical extent of $1.4\,\mathrm{kpc}$, as a more compact size would be inconsistent with the visual extinction measured by JWST. We note that, given the resolution of our data, dust of this extent would not be resolved across multiple ALMA beams. Combined with a fiducial yield of $0.1\,M_\odot$/SN, the AFM predicts a continuum flux density for GS-z14 of $F_{88} = 14.9\,\mu\mathrm{Jy}$. This is just below the sensitivity limit of our observations, which yield an upper limit of $F_{88} < 15.1\,\mu\mathrm{Jy\,beam}^{-1}$. \textcolor{black}{However, if the assumed dust yield is 0.5 dex lower ($0.1\,M_\odot$/SN), then the expected flux density would be $F_{88} \sim 8\,\mu\mathrm{Jy}$, well below the current limit.}  As such, a conclusive investigation of the dust content of GS-z14, as well as further testing of the AFM model predictions, requires deeper and high-resolution ALMA continuum observations. 



\begin{table}[ht]
\centering
\caption{Properties of GS-z14}
\vspace{-0.4cm}
\begin{tabular}{p{0.6\columnwidth} p{0.3\columnwidth}}
\hline
\textbf{Parameter} & \textbf{Value} \\ \hline
RA & 03:32:19.9049 \\ 
Dec & $-$27:51:20.265 \\ 
Redshift & $z=14.1793(7)$ \\ 
M$_{UV}$ & $-20.81\pm0.16$ \\ 
Stellar Mass ($\log(M_\odot)$) & $8.7_{-0.4}^{+0.5}$ \\
Star Formation Rate (M$_{\odot}$/yr) & $25_{-5}^{+6}$ \\
\oiii Luminosity ($10^8\,$L$_{\odot}$) & 2.1$\pm$0.5\\
FWHM \oiii (km s$^{-1}$) & 136$\pm$31 \\
Dynamical Mass (M$_{\odot}$ ($\mathrm{sin}\, i)^2$) & (1.0$\pm$0.5)$\times$10$^9$ \\
90-$\mu$m continuum flux ($\mu$Jy/beam) & $<15.1$  ($3\sigma$)\\
Dust Mass ($\log(M_\odot)$) & $<6.0$ \\ \hline
\end{tabular}
\begin{flushleft}
\footnotesize{\textit{Notes:} $M_\mathrm{UV}$ from \citet{carniani2024}; stellar mass and SFR from \citet{helton2024}.  Values have been corrected for a lensing magnification of 1.17$\times$ \citep{carniani2024}.}
\end{flushleft}
\end{table}

\section{Summary} \label{sec:summary}

We report the robust detection of a 6.6$\sigma$ [OIII]$_{88\mu m}\,$ line of JADES-GS-z14-0 at 223.528$\pm$0.009 GHz, providing us with a precise spectroscopic redshift measurement of $z=14.1793\pm0.0007$.  This represents a substantial jump in redshift over the previous high-redshift \oiii-detection from MACS1149-JD1 at $z=9.1096$ \citep{hashimoto2018}. The \oiii line was identified using data from an ALMA Cycle-10 DDT program (2023.A.00037.S, PI: Schouws) providing a spectral scan from 218.70 to 229.45 GHz (10.75 GHz baseline), covering the redshift range z=13.79 to 14.51.  The precision of the current redshift measurement represents $\gtrsim180$$\times$ gain over the prior redshift measurement of $z=14.32_{-0.20}^{+0.08}$ from NIRSpec.

The redshift we find for the source is consistent with the redshift \citet{carniani2024} derive ($z=14.178\pm0.013$) based on their tentative 3.6$\sigma$ detection of CIII]$_{1907,1909}$ doublet at 2.89$\mu$m, providing strong evidence the earlier line detection is real.  As such, we now have multiple line detections of GS-z14-0, CIII]$_{1907,1909}$ (3.6$\sigma$) with JWST and \oiii$\,$ (6.6$\sigma$) with ALMA, providing rather definitive evidence for the robustness of the redshift determination of GS-z14-0.  Of note, the detection of \oiii$\,$ with ALMA was achieved with less integration time (2.8 hours) than was required for the 3.6$\sigma$ tentative detection of CIII]$_{1907,1909}$ with JWST (9.3 hours), providing a rather dramatic illustration of the discovery potential of ALMA.

We find no detection of the dust continuum from JADES-GS-z14-0 based on the DDT observations, with a $<$ $3\sigma$ upper limit of $<$15.1$\,\mu$Jy beam$^{-1}$. This suggests a low dust-to-stellar mass ratio of $M_\mathrm{d}/M_\star < 1.9 \times 10^{-3}$, consistent with supernova dust production yields $y_d < 0.24\,M_\odot$/SN. 

Combining a previous JWST/MIRI photometric measurement of the  \oojwst~ and H$\beta$ lines with {\sc Cloudy} models, we find GS-z14 to be surprisingly metal-enriched ($Z\sim 0.05$-0.2 Z$_{\odot}$) a mere 300 Myr after the Big Bang, with moderate to low density $\log(N) < 3.0\,\textrm{cm}^{-3}$. 

Thanks to the precise spectroscopic redshift measurement and \oiii line detection we now have for GS-14-0 using just 2.8 hours, it is clear that additional follow-up of GS-z14-0 with ALMA would be highly fruitful and should include (1) higher spatial resolution observations of \oiii to improve constraints on dynamical masses and state of GS-z14-0 and deeper constraints on the dust continuum, (2) observations of the [CII]$_{158\mu m}$ line in band 4, lying just 0.2 GHz above the low-frequency boundary, to better probe the ionization parameter $U$ and build-up of metals, and (3) JWST/MIRI observations of \oojwst\, and \Hb\, to place tighter constraints on the electron density and metallicity of GS-z14-0.


\begin{acknowledgments}
\subsection*{Acknowledgments} 
This paper makes use of the following ALMA data: ADS/JAO.ALMA 2023.A.00037.S. ALMA is a partnership of ESO (representing its member states), NSF (USA) and NINS (Japan), together with NRC (Canada), MOST and ASIAA (Taiwan), and KASI (Republic of Korea), in cooperation with the Republic of Chile. The Joint ALMA Observatory is operated by ESO, AUI/NRAO and NAOJ. We are greatly appreciative to our ALMA program coordinator Violette Impellizzeri for support with our ALMA program and Allegro, the European ALMA Regional Center node in the Netherlands. We are grateful to J. Witstok for providing the {\sc Cloudy} models from \citet{Witstok_2022} for the analysis conducted in this paper. We thank M. Kohandel and A. Pallottini for simulated data analysis. 
This work was supported by NAOJ ALMA Scientific Research Grant Code 2021-19A (HA). AF acknowledges support from the ERC Advanced Grant INTERSTELLAR H2020/740120 and support from the grant NSF PHY-2309135 to the Kavli Institute for Theoretical Physics (KITP). 
 
\end{acknowledgments}


\bibliography{sample631}{}

\begin{thebibliography}{}
\expandafter\ifx\csname natexlab\endcsname\relax\def\natexlab#1{#1}\fi
\providecommand{\url}[1]{\href{#1}{#1}}
\providecommand{\dodoi}[1]{doi:~\href{http://doi.org/#1}{\nolinkurl{#1}}}
\providecommand{\doeprint}[1]{\href{http://ascl.net/#1}{\nolinkurl{http://ascl.net/#1}}}
\providecommand{\doarXiv}[1]{\href{https://arxiv.org/abs/#1}{\nolinkurl{https://arxiv.org/abs/#1}}}

\bibitem[{{Akins} {et~al.}(2022){Akins}, {Fujimoto}, {Finlator}, {Watson}, {Knudsen}, {Richard}, {Bakx}, {Hashimoto}, {Inoue}, {Matsuo}, {Micha{\l}owski}, \& {Tamura}}]{akins2022}
{Akins}, H.~B., {Fujimoto}, S., {Finlator}, K., {et~al.} 2022, \apj, 934, 64, \dodoi{10.3847/1538-4357/ac795b}

\bibitem[{{Algera} {et~al.}(2023){Algera}, {Inami}, {Oesch}, {Sommovigo}, {Bouwens}, {Topping}, {Schouws}, {Stefanon}, {Stark}, {Aravena}, {Barrufet}, {da Cunha}, {Dayal}, {Endsley}, {Ferrara}, {Fudamoto}, {Gonzalez}, {Graziani}, {Hodge}, {Hygate}, {de Looze}, {Nanayakkara}, {Schneider}, \& {van der Werf}}]{algera2023}
{Algera}, H. S.~B., {Inami}, H., {Oesch}, P.~A., {et~al.} 2023, \mnras, 518, 6142, \dodoi{10.1093/mnras/stac3195}

\bibitem[{{Algera} {et~al.}(2024){Algera}, {Inami}, {Sommovigo}, {Fudamoto}, {Schneider}, {Graziani}, {Dayal}, {Bouwens}, {Aravena}, {da Cunha}, {Ferrara}, {Hygate}, {van Leeuwen}, {De Looze}, {Palla}, {Pallottini}, {Smit}, {Stefanon}, {Topping}, \& {van der Werf}}]{algera2024}
{Algera}, H. S.~B., {Inami}, H., {Sommovigo}, L., {et~al.} 2024, \mnras, 527, 6867, \dodoi{10.1093/mnras/stad3111}

\bibitem[{{ALMA Technical Handbook}(2024)}]{ALMA_Tech_Handbook_C10}
{ALMA Technical Handbook}. 2024, ALMA Cycle 10 Technical Handbook, ALMA Observatory

\bibitem[{{Arata} {et~al.}(2020){Arata}, {Yajima}, {Nagamine}, {Abe}, \& {Khochfar}}]{arata2020}
{Arata}, S., {Yajima}, H., {Nagamine}, K., {Abe}, M., \& {Khochfar}, S. 2020, \mnras, 498, 5541, \dodoi{10.1093/mnras/staa2809}

\bibitem[{{Asplund} {et~al.}(2009){Asplund}, {Grevesse}, {Sauval}, \& {Scott}}]{asplund2009}
{Asplund}, M., {Grevesse}, N., {Sauval}, A.~J., \& {Scott}, P. 2009, \araa, 47, 481, \dodoi{10.1146/annurev.astro.46.060407.145222}

\bibitem[{{Bakx} {et~al.}(2020){Bakx}, {Tamura}, {Hashimoto}, {Inoue}, {Lee}, {Mawatari}, {Ota}, {Umehata}, {Zackrisson}, {Hatsukade}, {Kohno}, {Matsuda}, {Matsuo}, {Okamoto}, {Shibuya}, {Shimizu}, {Taniguchi}, \& {Yoshida}}]{bakx2020}
{Bakx}, T. J.~L.~C., {Tamura}, Y., {Hashimoto}, T., {et~al.} 2020, \mnras, 493, 4294, \dodoi{10.1093/mnras/staa509}

\bibitem[{{Bakx} {et~al.}(2023){Bakx}, {Zavala}, {Mitsuhashi}, {Treu}, {Fontana}, {Tadaki}, {Casey}, {Castellano}, {Glazebrook}, {Hagimoto}, {Ikeda}, {Jones}, {Leethochawalit}, {Mason}, {Morishita}, {Nanayakkara}, {Pentericci}, {Roberts-Borsani}, {Santini}, {Serjeant}, {Tamura}, {Trenti}, \& {Vanzella}}]{bakx2023}
{Bakx}, T. J.~L.~C., {Zavala}, J.~A., {Mitsuhashi}, I., {et~al.} 2023, \mnras, 519, 5076, \dodoi{10.1093/mnras/stac3723}

\bibitem[{{B{\'e}thermin} {et~al.}(2020){B{\'e}thermin}, {Fudamoto}, {Ginolfi}, {Loiacono}, {Khusanova}, {Capak}, {Cassata}, {Faisst}, {Le F{\`e}vre}, {Schaerer}, {Silverman}, {Yan}, {Amorin}, {Bardelli}, {Boquien}, {Cimatti}, {Davidzon}, {Dessauges-Zavadsky}, {Fujimoto}, {Gruppioni}, {Hathi}, {Ibar}, {Jones}, {Koekemoer}, {Lagache}, {Lemaux}, {Moreau}, {Oesch}, {Pozzi}, {Riechers}, {Talia}, {Toft}, {Vallini}, {Vergani}, {Zamorani}, \& {Zucca}}]{bethermin2020}
{B{\'e}thermin}, M., {Fudamoto}, Y., {Ginolfi}, M., {et~al.} 2020, \aap, 643, A2, \dodoi{10.1051/0004-6361/202037649}

\bibitem[{{Bouwens} {et~al.}(2023){Bouwens}, {Illingworth}, {Oesch}, {Stefanon}, {Naidu}, {van Leeuwen}, \& {Magee}}]{bouwens2023}
{Bouwens}, R., {Illingworth}, G., {Oesch}, P., {et~al.} 2023, \mnras, 523, 1009, \dodoi{10.1093/mnras/stad1014}

\bibitem[{{Bouwens} {et~al.}(2022){Bouwens}, {Smit}, {Schouws}, {Stefanon}, {Bowler}, {Endsley}, {Gonzalez}, {Inami}, {Stark}, {Oesch}, {Hodge}, {Aravena}, {da Cunha}, {Dayal}, {Looze}, {Ferrara}, {Fudamoto}, {Graziani}, {Li}, {Nanayakkara}, {Pallottini}, {Schneider}, {Sommovigo}, {Topping}, {van der Werf}, {Algera}, {Barrufet}, {Hygate}, {Labb{\'e}}, {Riechers}, \& {Witstok}}]{bouwens2022}
{Bouwens}, R.~J., {Smit}, R., {Schouws}, S., {et~al.} 2022, \apj, 931, 160, \dodoi{10.3847/1538-4357/ac5a4a}

\bibitem[{Bunker {et~al.}(2023)Bunker, Saxena, Cameron, Willott, Curtis-Lake, Jakobsen, Carniani, Smit, Maiolino, Witstok, Curti, D’Eugenio, Jones, Ferruit, Arribas, Charlot, Chevallard, Giardino, de~Graaff, Looser, Lützgendorf, Maseda, Rawle, Rix, Del~Pino, Alberts, Egami, Eisenstein, Endsley, Hainline, Hausen, Johnson, Rieke, Rieke, Robertson, Shivaei, Stark, Sun, Tacchella, Tang, Williams, Willmer, Baker, Baum, Bhatawdekar, Bowler, Boyett, Chen, Circosta, Helton, Ji, Kumari, Lyu, Nelson, Parlanti, Perna, Sandles, Scholtz, Suess, Topping, Übler, Wallace, \& Whitler}]{Bunker_2023}
Bunker, A.~J., Saxena, A., Cameron, A.~J., {et~al.} 2023, Astronomy \& Astrophysics, 677, A88, \dodoi{10.1051/0004-6361/202346159}

\bibitem[{{Bunker} {et~al.}(2023){Bunker}, {Saxena}, {Cameron}, {Willott}, {Curtis-Lake}, {Jakobsen}, {Carniani}, {Smit}, {Maiolino}, {Witstok}, {Curti}, {D'Eugenio}, {Jones}, {Ferruit}, {Arribas}, {Charlot}, {Chevallard}, {Giardino}, {de Graaff}, {Looser}, {L{\"u}tzgendorf}, {Maseda}, {Rawle}, {Rix}, {Del Pino}, {Alberts}, {Egami}, {Eisenstein}, {Endsley}, {Hainline}, {Hausen}, {Johnson}, {Rieke}, {Rieke}, {Robertson}, {Shivaei}, {Stark}, {Sun}, {Tacchella}, {Tang}, {Williams}, {Willmer}, {Baker}, {Baum}, {Bhatawdekar}, {Bowler}, {Boyett}, {Chen}, {Circosta}, {Helton}, {Ji}, {Kumari}, {Lyu}, {Nelson}, {Parlanti}, {Perna}, {Sandles}, {Scholtz}, {Suess}, {Topping}, {{\"U}bler}, {Wallace}, \& {Whitler}}]{bunker2023}
{Bunker}, A.~J., {Saxena}, A., {Cameron}, A.~J., {et~al.} 2023, \aap, 677, A88, \dodoi{10.1051/0004-6361/202346159}

\bibitem[{{Carniani} {et~al.}(2020){Carniani}, {Ferrara}, {Maiolino}, {Castellano}, {Gallerani}, {Fontana}, {Kohandel}, {Lupi}, {Pallottini}, {Pentericci}, {Vallini}, \& {Vanzella}}]{carniani2020}
{Carniani}, S., {Ferrara}, A., {Maiolino}, R., {et~al.} 2020, \mnras, 499, 5136, \dodoi{10.1093/mnras/staa3178}

\bibitem[{{Carniani} {et~al.}(2024){Carniani}, {Hainline}, {D'Eugenio}, {Eisenstein}, {Jakobsen}, {Witstok}, {Johnson}, {Chevallard}, {Maiolino}, {Helton}, {Willott}, {Robertson}, {Alberts}, {Arribas}, {Baker}, {Bhatawdekar}, {Boyett}, {Bunker}, {Cameron}, {Cargile}, {Charlot}, {Curti}, {Curtis-Lake}, {Egami}, {Giardino}, {Isaak}, {Ji}, {Jones}, {Kumari}, {Maseda}, {Parlanti}, {P{\'e}rez-Gonz{\'a}lez}, {Rawle}, {Rieke}, {Rieke}, {Del Pino}, {Saxena}, {Scholtz}, {Smit}, {Sun}, {Tacchella}, {{\"U}bler}, {Venturi}, {Williams}, \& {Willmer}}]{carniani2024}
{Carniani}, S., {Hainline}, K., {D'Eugenio}, F., {et~al.} 2024, \nat, 633, 318, \dodoi{10.1038/s41586-024-07860-9}

\bibitem[{{Castellano} {et~al.}(2022){Castellano}, {Fontana}, {Treu}, {Santini}, {Merlin}, {Leethochawalit}, {Trenti}, {Vanzella}, {Mestric}, {Bonchi}, {Belfiori}, {Nonino}, {Paris}, {Polenta}, {Roberts-Borsani}, {Boyett}, {Brada{\v{c}}}, {Calabr{\`o}}, {Glazebrook}, {Grillo}, {Mascia}, {Mason}, {Mercurio}, {Morishita}, {Nanayakkara}, {Pentericci}, {Rosati}, {Vulcani}, {Wang}, \& {Yang}}]{castellano2022}
{Castellano}, M., {Fontana}, A., {Treu}, T., {et~al.} 2022, \apjl, 938, L15, \dodoi{10.3847/2041-8213/ac94d0}

\bibitem[{{Castellano} {et~al.}(2024){Castellano}, {Napolitano}, {Fontana}, {Roberts-Borsani}, {Treu}, {Vanzella}, {Zavala}, {Arrabal Haro}, {Calabr{\`o}}, {Llerena}, {Mascia}, {Merlin}, {Paris}, {Pentericci}, {Santini}, {Bakx}, {Bergamini}, {Cupani}, {Dickinson}, {Filippenko}, {Glazebrook}, {Grillo}, {Kelly}, {Malkan}, {Mason}, {Morishita}, {Nanayakkara}, {Rosati}, {Sani}, {Wang}, \& {Yoon}}]{castellano2024}
{Castellano}, M., {Napolitano}, L., {Fontana}, A., {et~al.} 2024, \apj, 972, 143, \dodoi{10.3847/1538-4357/ad5f88}

\bibitem[{{Chabrier}(2003)}]{chabrier2003}
{Chabrier}, G. 2003, \pasp, 115, 763, \dodoi{10.1086/376392}

\bibitem[{{Cormier} {et~al.}(2015){Cormier}, {Madden}, {Lebouteiller}, {Abel}, {Hony}, {Galliano}, {R{\'e}my-Ruyer}, {Bigiel}, {Baes}, {Boselli}, {Chevance}, {Cooray}, {De Looze}, {Doublier}, {Galametz}, {Hughes}, {Karczewski}, {Lee}, {Lu}, \& {Spinoglio}}]{Cormier2015}
{Cormier}, D., {Madden}, S.~C., {Lebouteiller}, V., {et~al.} 2015, \aap, 578, A53, \dodoi{10.1051/0004-6361/201425207}

\bibitem[{{Curti} {et~al.}(2024){Curti}, {Maiolino}, {Curtis-Lake}, {Chevallard}, {Carniani}, {D'Eugenio}, {Looser}, {Scholtz}, {Charlot}, {Cameron}, {{\"U}bler}, {Witstok}, {Boyett}, {Laseter}, {Sandles}, {Arribas}, {Bunker}, {Giardino}, {Maseda}, {Rawle}, {Rodr{\'\i}guez Del Pino}, {Smit}, {Willott}, {Eisenstein}, {Hausen}, {Johnson}, {Rieke}, {Robertson}, {Tacchella}, {Williams}, {Willmer}, {Baker}, {Bhatawdekar}, {Egami}, {Helton}, {Ji}, {Kumari}, {Perna}, {Shivaei}, \& {Sun}}]{curti2024}
{Curti}, M., {Maiolino}, R., {Curtis-Lake}, E., {et~al.} 2024, \aap, 684, A75, \dodoi{10.1051/0004-6361/202346698}

\bibitem[{{Curtis-Lake} {et~al.}(2023){Curtis-Lake}, {Carniani}, {Cameron}, {Charlot}, {Jakobsen}, {Maiolino}, {Bunker}, {Witstok}, {Smit}, {Chevallard}, {Willott}, {Ferruit}, {Arribas}, {Bonaventura}, {Curti}, {D'Eugenio}, {Franx}, {Giardino}, {Looser}, {L{\"u}tzgendorf}, {Maseda}, {Rawle}, {Rix}, {Rodr{\'\i}guez del Pino}, {{\"U}bler}, {Sirianni}, {Dressler}, {Egami}, {Eisenstein}, {Endsley}, {Hainline}, {Hausen}, {Johnson}, {Rieke}, {Robertson}, {Shivaei}, {Stark}, {Tacchella}, {Williams}, {Willmer}, {Bhatawdekar}, {Bowler}, {Boyett}, {Chen}, {de Graaff}, {Helton}, {Hviding}, {Jones}, {Kumari}, {Lyu}, {Nelson}, {Perna}, {Sandles}, {Saxena}, {Suess}, {Sun}, {Topping}, {Wallace}, \& {Whitler}}]{Curtis-Lake_2022}
{Curtis-Lake}, E., {Carniani}, S., {Cameron}, A., {et~al.} 2023, Nature Astronomy, 7, 622, \dodoi{10.1038/s41550-023-01918-w}

\bibitem[{Dayal {et~al.}(2022)Dayal, Ferrara, Sommovigo, Bouwens, Oesch, Smit, Gonzalez, Schouws, Stefanon, Kobayashi, Bremer, Algera, Aravena, Bowler, da Cunha, Fudamoto, Graziani, Hodge, Inami, De Looze, Pallottini, Riechers, Schneider, Stark, \& Endsley}]{Dayal_2022}
Dayal, P., Ferrara, A., Sommovigo, L., {et~al.} 2022, Monthly Notices of the Royal Astronomical Society, 512, 989–1002, \dodoi{10.1093/mnras/stac537}

\bibitem[{{Dayal} {et~al.}(2022){Dayal}, {Ferrara}, {Sommovigo}, {Bouwens}, {Oesch}, {Smit}, {Gonzalez}, {Schouws}, {Stefanon}, {Kobayashi}, {Bremer}, {Algera}, {Aravena}, {Bowler}, {da Cunha}, {Fudamoto}, {Graziani}, {Hodge}, {Inami}, {De Looze}, {Pallottini}, {Riechers}, {Schneider}, {Stark}, \& {Endsley}}]{dayal2022}
{Dayal}, P., {Ferrara}, A., {Sommovigo}, L., {et~al.} 2022, \mnras, 512, 989, \dodoi{10.1093/mnras/stac537}

\bibitem[{{De Looze} {et~al.}(2014){De Looze}, {Cormier}, {Lebouteiller}, {Madden}, {Baes}, {Bendo}, {Boquien}, {Boselli}, {Clements}, {Cortese}, {Cooray}, {Galametz}, {Galliano}, {Graci{\'a}-Carpio}, {Isaak}, {Karczewski}, {Parkin}, {Pellegrini}, {R{\'e}my-Ruyer}, {Spinoglio}, {Smith}, \& {Sturm}}]{delooze2014}
{De Looze}, I., {Cormier}, D., {Lebouteiller}, V., {et~al.} 2014, \aap, 568, A62, \dodoi{10.1051/0004-6361/201322489}

\bibitem[{{De Looze} {et~al.}(2020){De Looze}, {Lamperti}, {Saintonge}, {Rela{\~n}o}, {Smith}, {Clark}, {Wilson}, {Decleir}, {Jones}, {Kennicutt}, {Accurso}, {Brinks}, {Bureau}, {Cigan}, {Clements}, {De Vis}, {Fanciullo}, {Gao}, {Gear}, {Ho}, {Hwang}, {Micha{\l}owski}, {Lee}, {Li}, {Lin}, {Liu}, {Lomaeva}, {Pan}, {Sargent}, {Williams}, {Xiao}, \& {Zhu}}]{delooze2020}
{De Looze}, I., {Lamperti}, I., {Saintonge}, A., {et~al.} 2020, \mnras, 496, 3668, \dodoi{10.1093/mnras/staa1496}

\bibitem[{Donnan {et~al.}(2022)Donnan, McLeod, Dunlop, McLure, Carnall, Begley, Cullen, Hamadouche, Bowler, Magee, McCracken, Milvang-Jensen, Moneti, \& Targett}]{Donnan_2022}
Donnan, C.~T., McLeod, D.~J., Dunlop, J.~S., {et~al.} 2022, Monthly Notices of the Royal Astronomical Society, 518, 6011–6040, \dodoi{10.1093/mnras/stac3472}

\bibitem[{{Eisenstein} {et~al.}(2023{\natexlab{a}}){Eisenstein}, {Willott}, {Alberts}, {Arribas}, {Bonaventura}, {Bunker}, {Cameron}, {Carniani}, {Charlot}, {Curtis-Lake}, {D'Eugenio}, {Endsley}, {Ferruit}, {Giardino}, {Hainline}, {Hausen}, {Jakobsen}, {Johnson}, {Maiolino}, {Rieke}, {Rieke}, {Rix}, {Robertson}, {Stark}, {Tacchella}, {Williams}, {Willmer}, {Baker}, {Baum}, {Bhatawdekar}, {Boyett}, {Chen}, {Chevallard}, {Circosta}, {Curti}, {Danhaive}, {DeCoursey}, {de Graaff}, {Dressler}, {Egami}, {Helton}, {Hviding}, {Ji}, {Jones}, {Kumari}, {L{\"u}tzgendorf}, {Laseter}, {Looser}, {Lyu}, {Maseda}, {Nelson}, {Parlanti}, {Perna}, {Pusk{\'a}s}, {Rawle}, {Rodr{\'\i}guez Del Pino}, {Sandles}, {Saxena}, {Scholtz}, {Sharpe}, {Shivaei}, {Silcock}, {Simmonds}, {Skarbinski}, {Smit}, {Stone}, {Suess}, {Sun}, {Tang}, {Topping}, {{\"U}bler}, {Villanueva}, {Wallace}, {Whitler}, {Witstok}, \& {Woodrum}}]{Eisenstein_2023}
{Eisenstein}, D.~J., {Willott}, C., {Alberts}, S., {et~al.} 2023{\natexlab{a}}, arXiv e-prints, arXiv:2306.02465, \dodoi{10.48550/arXiv.2306.02465}

\bibitem[{{Eisenstein} {et~al.}(2023{\natexlab{b}}){Eisenstein}, {Johnson}, {Robertson}, {Tacchella}, {Hainline}, {Jakobsen}, {Maiolino}, {Bonaventura}, {Bunker}, {Cameron}, {Cargile}, {Curtis-Lake}, {Hausen}, {Pusk{\'a}s}, {Rieke}, {Sun}, {Willmer}, {Willott}, {Alberts}, {Arribas}, {Baker}, {Baum}, {Bhatawdekar}, {Carniani}, {Charlot}, {Chen}, {Chevallard}, {Curti}, {DeCoursey}, {D'Eugenio}, {de Graaff}, {Egami}, {Helton}, {Ji}, {Jones}, {Kumari}, {L{\"u}tzgendorf}, {Laseter}, {Looser}, {Lyu}, {Maseda}, {Nelson}, {Parlanti}, {Rauscher}, {Rawle}, {Rieke}, {Rix}, {Rujopakarn}, {Sandles}, {Saxena}, {Scholtz}, {Sharpe}, {Shivaei}, {Simmonds}, {Smit}, {Topping}, {{\"U}bler}, {Venturi}, {Williams}, {Witstok}, \& {Woodrum}}]{Eisenstein_2023_JOF}
{Eisenstein}, D.~J., {Johnson}, B.~D., {Robertson}, B., {et~al.} 2023{\natexlab{b}}, arXiv e-prints, arXiv:2310.12340, \dodoi{10.48550/arXiv.2310.12340}

\bibitem[{Faisst {et~al.}(2020)Faisst, Fudamoto, Oesch, Scoville, Riechers, Pavesi, \& Capak}]{faisst2020_tdust}
Faisst, A.~L., Fudamoto, Y., Oesch, P.~A., {et~al.} 2020, Monthly Notices of the Royal Astronomical Society, 498, 4192, \dodoi{10.1093/mnras/staa2545}

\bibitem[{{Faisst} {et~al.}(2020){Faisst}, {Schaerer}, {Lemaux}, {Oesch}, {Fudamoto}, {Cassata}, {B{\'e}thermin}, {Capak}, {Le F{\`e}vre}, {Silverman}, {Yan}, {Ginolfi}, {Koekemoer}, {Morselli}, {Amor{\'\i}n}, {Bardelli}, {Boquien}, {Brammer}, {Cimatti}, {Dessauges-Zavadsky}, {Fujimoto}, {Gruppioni}, {Hathi}, {Hemmati}, {Ibar}, {Jones}, {Khusanova}, {Loiacono}, {Pozzi}, {Talia}, {Tasca}, {Riechers}, {Rodighiero}, {Romano}, {Scoville}, {Toft}, {Vallini}, {Vergani}, {Zamorani}, \& {Zucca}}]{faisst2020}
{Faisst}, A.~L., {Schaerer}, D., {Lemaux}, B.~C., {et~al.} 2020, \apjs, 247, 61, \dodoi{10.3847/1538-4365/ab7ccd}

\bibitem[{{Ferland} {et~al.}(2017){Ferland}, {Chatzikos}, {Guzm{\'a}n}, {Lykins}, {van Hoof}, {Williams}, {Abel}, {Badnell}, {Keenan}, {Porter}, \& {Stancil}}]{ferland2017}
{Ferland}, G.~J., {Chatzikos}, M., {Guzm{\'a}n}, F., {et~al.} 2017, \rmxaa, 53, 385.
\newblock \doarXiv{1705.10877}

\bibitem[{{Ferrara}(2024)}]{Ferrara2024a}
{Ferrara}, A. 2024, \aap, 689, A310, \dodoi{10.1051/0004-6361/202450944}

\bibitem[{Ferrara {et~al.}(2024)Ferrara, Carniani, di~Mascia, Bouwens, Oesch, \& Schouws}]{Ferrara2024b}
Ferrara, A., Carniani, S., di~Mascia, F., {et~al.} 2024.
\newblock \doarXiv{2409.17223}

\bibitem[{{Ferrara} {et~al.}(2016){Ferrara}, {Viti}, \& {Ceccarelli}}]{ferrara2016}
{Ferrara}, A., {Viti}, S., \& {Ceccarelli}, C. 2016, \mnras, 463, L112, \dodoi{10.1093/mnrasl/slw165}

\bibitem[{{Finkelstein} {et~al.}(2024){Finkelstein}, {Leung}, {Bagley}, {Dickinson}, {Ferguson}, {Papovich}, {Akins}, {Arrabal Haro}, {Dav{\'e}}, {Dekel}, {Kartaltepe}, {Kocevski}, {Koekemoer}, {Pirzkal}, {Somerville}, {Yung}, {Amor{\'\i}n}, {Backhaus}, {Behroozi}, {Bisigello}, {Bromm}, {Casey}, {Ch{\'a}vez Ortiz}, {Cheng}, {Chworowsky}, {Cleri}, {Cooper}, {Davis}, {de la Vega}, {Elbaz}, {Franco}, {Fontana}, {Fujimoto}, {Giavalisco}, {Grogin}, {Holwerda}, {Huertas-Company}, {Hirschmann}, {Iyer}, {Jogee}, {Jung}, {Larson}, {Lucas}, {Mobasher}, {Morales}, {Morley}, {Mukherjee}, {P{\'e}rez-Gonz{\'a}lez}, {Ravindranath}, {Rodighiero}, {Rowland}, {Tacchella}, {Taylor}, {Trump}, \& {Wilkins}}]{Finkelstein_2023}
{Finkelstein}, S.~L., {Leung}, G. C.~K., {Bagley}, M.~B., {et~al.} 2024, \apjl, 969, L2, \dodoi{10.3847/2041-8213/ad4495}

\bibitem[{{Fudamoto} {et~al.}(2021){Fudamoto}, {Oesch}, {Schouws}, {Stefanon}, {Smit}, {Bouwens}, {Bowler}, {Endsley}, {Gonzalez}, {Inami}, {Labbe}, {Stark}, {Aravena}, {Barrufet}, {da Cunha}, {Dayal}, {Ferrara}, {Graziani}, {Hodge}, {Hutter}, {Li}, {De Looze}, {Nanayakkara}, {Pallottini}, {Riechers}, {Schneider}, {Ucci}, {van der Werf}, \& {White}}]{fudamoto2021}
{Fudamoto}, Y., {Oesch}, P.~A., {Schouws}, S., {et~al.} 2021, \nat, 597, 489, \dodoi{10.1038/s41586-021-03846-z}

\bibitem[{{Fujimoto} {et~al.}(2023){Fujimoto}, {Finkelstein}, {Burgarella}, {Carilli}, {Buat}, {Casey}, {Ciesla}, {Tacchella}, {Zavala}, {Brammer}, {Fudamoto}, {Ouchi}, {Valentino}, {Cooper}, {Dickinson}, {Franco}, {Giavalisco}, {Hutchison}, {Kartaltepe}, {Koekemoer}, {Kojima}, {Larson}, {Murphy}, {Papovich}, {P{\'e}rez-Gonz{\'a}lez}, {Somerville}, {Yoon}, {Wilkins}, {Akins}, {Amor{\'\i}n}, {Arrabal Haro}, {Bagley}, {Chworowsky}, {Cleri}, {Cooper}, {Costantin}, {Daddi}, {Ferguson}, {Grogin}, {Jim{\'e}nez-Andrade}, {Juneau}, {Kirkpatrick}, {Kocevski}, {Le Bail}, {Long}, {Lucas}, {Magnelli}, {McKinney}, {Rose}, {Seill{\'e}}, {Simons}, {Weiner}, \& {Yung}}]{Fujimoto_2022}
{Fujimoto}, S., {Finkelstein}, S.~L., {Burgarella}, D., {et~al.} 2023, \apj, 955, 130, \dodoi{10.3847/1538-4357/aceb67}

\bibitem[{{Fujimoto} {et~al.}(2024){Fujimoto}, {Ouchi}, {Nakajima}, {Harikane}, {Isobe}, {Brammer}, {Oguri}, {Gim{\'e}nez-Arteaga}, {Heintz}, {Kokorev}, {Bauer}, {Ferrara}, {Kojima}, {Lagos}, {Laura}, {Schaerer}, {Shimasaku}, {Hatsukade}, {Kohno}, {Sun}, {Valentino}, {Watson}, {Fudamoto}, {Inoue}, {Gonz{\'a}lez-L{\'o}pez}, {Koekemoer}, {Knudsen}, {Lee}, {Magdis}, {Richard}, {Strait}, {Sugahara}, {Tamura}, {Toft}, {Umehata}, \& {Walth}}]{fujimoto2024a}
{Fujimoto}, S., {Ouchi}, M., {Nakajima}, K., {et~al.} 2024, \apj, 964, 146, \dodoi{10.3847/1538-4357/ad235c}

\bibitem[{Furlanetto \& Mirocha(2023)}]{Furlanetto_2023}
Furlanetto, S.~R., \& Mirocha, J. 2023, Monthly Notices of the Royal Astronomical Society, 523, 5274–5279, \dodoi{10.1093/mnras/stad1799}

\bibitem[{{Galliano} {et~al.}(2021){Galliano}, {Nersesian}, {Bianchi}, {De Looze}, {Roychowdhury}, {Baes}, {Casasola}, {Cassar{\'a}}, {Dobbels}, {Fritz}, {Galametz}, {Jones}, {Madden}, {Mosenkov}, {Xilouris}, \& {Ysard}}]{galliano2021}
{Galliano}, F., {Nersesian}, A., {Bianchi}, S., {et~al.} 2021, \aap, 649, A18, \dodoi{10.1051/0004-6361/202039701}

\bibitem[{Genzel {et~al.}(2010)Genzel, Tacconi, Gracia-Carpio, Sternberg, Cooper, Shapiro, Bolatto, Bouché, Bournaud, Burkert, Combes, Comerford, Cox, Davis, Schreiber, Garcia-Burillo, Lutz, Naab, Neri, Omont, Shapley, \& Weiner}]{Genzel_2010}
Genzel, R., Tacconi, L.~J., Gracia-Carpio, J., {et~al.} 2010, Monthly Notices of the Royal Astronomical Society, 407, 2091–2108, \dodoi{10.1111/j.1365-2966.2010.16969.x}

\bibitem[{{Hainline} {et~al.}(2024){Hainline}, {Johnson}, {Robertson}, {Tacchella}, {Helton}, {Sun}, {Eisenstein}, {Simmonds}, {Topping}, {Whitler}, {Willmer}, {Rieke}, {Suess}, {Hviding}, {Cameron}, {Alberts}, {Baker}, {Baum}, {Bhatawdekar}, {Bonaventura}, {Boyett}, {Bunker}, {Carniani}, {Charlot}, {Chevallard}, {Chen}, {Curti}, {Curtis-Lake}, {D'Eugenio}, {Egami}, {Endsley}, {Hausen}, {Ji}, {Looser}, {Lyu}, {Maiolino}, {Nelson}, {Pusk{\'a}s}, {Rawle}, {Sandles}, {Saxena}, {Smit}, {Stark}, {Williams}, {Willott}, \& {Witstok}}]{Hainline_2023}
{Hainline}, K.~N., {Johnson}, B.~D., {Robertson}, B., {et~al.} 2024, \apj, 964, 71, \dodoi{10.3847/1538-4357/ad1ee4}

\bibitem[{{Harikane} {et~al.}(2020){Harikane}, {Ouchi}, {Inoue}, {Matsuoka}, {Tamura}, {Bakx}, {Fujimoto}, {Moriwaki}, {Ono}, {Nagao}, {Tadaki}, {Kojima}, {Shibuya}, {Egami}, {Ferrara}, {Gallerani}, {Hashimoto}, {Kohno}, {Matsuda}, {Matsuo}, {Pallottini}, {Sugahara}, \& {Vallini}}]{harikane2020}
{Harikane}, Y., {Ouchi}, M., {Inoue}, A.~K., {et~al.} 2020, \apj, 896, 93, \dodoi{10.3847/1538-4357/ab94bd}

\bibitem[{{Harikane} {et~al.}(2023){Harikane}, {Ouchi}, {Oguri}, {Ono}, {Nakajima}, {Isobe}, {Umeda}, {Mawatari}, \& {Zhang}}]{harikane2022}
{Harikane}, Y., {Ouchi}, M., {Oguri}, M., {et~al.} 2023, \apjs, 265, 5, \dodoi{10.3847/1538-4365/acaaa9}

\bibitem[{Harikane {et~al.}(2023)Harikane, Ouchi, Oguri, Ono, Nakajima, Isobe, Umeda, Mawatari, \& Zhang}]{Harikane_2023}
Harikane, Y., Ouchi, M., Oguri, M., {et~al.} 2023, The Astrophysical Journal Supplement Series, 265, 5, \dodoi{10.3847/1538-4365/acaaa9}

\bibitem[{{Harikane} {et~al.}(2024){Harikane}, {Inoue}, {Ellis}, {Ouchi}, {Nakazato}, {Yoshida}, {Ono}, {Sun}, {Sato}, {Fujimoto}, {Kashikawa}, {McLeod}, {Perez-Gonzalez}, {Sawicki}, {Sugahara}, {Xu}, {Yamanaka}, {Carnall}, {Cullen}, {Dunlop}, {Egami}, {Grogin}, {Isobe}, {Koekemoer}, {Laporte}, {Lee}, {Magee}, {Matsuo}, {Matsuoka}, {Mawatari}, {Nakajima}, {Nakane}, {Tamura}, {Umeda}, \& {Yanagisawa}}]{harikane2024}
{Harikane}, Y., {Inoue}, A.~K., {Ellis}, R.~S., {et~al.} 2024, arXiv e-prints, arXiv:2406.18352, \dodoi{10.48550/arXiv.2406.18352}

\bibitem[{{Hashimoto} {et~al.}(2018){Hashimoto}, {Laporte}, {Mawatari}, {Ellis}, {Inoue}, {Zackrisson}, {Roberts-Borsani}, {Zheng}, {Tamura}, {Bauer}, {Fletcher}, {Harikane}, {Hatsukade}, {Hayatsu}, {Matsuda}, {Matsuo}, {Okamoto}, {Ouchi}, {Pell{\'o}}, {Rydberg}, {Shimizu}, {Taniguchi}, {Umehata}, \& {Yoshida}}]{hashimoto2018}
{Hashimoto}, T., {Laporte}, N., {Mawatari}, K., {et~al.} 2018, \nat, 557, 392, \dodoi{10.1038/s41586-018-0117-z}

\bibitem[{{Heintz, K. E.} {et~al.}(2025){Heintz, K. E.}, {Brammer, G. B.}, {Watson, D.}, {Oesch, P. A.}, {Keating, L. C.}, {Hayes, M. J.}, {Abdurro’uf}, {Arellano-Córdova, K. Z.}, {Carnall, A. C.}, {Christiansen, C. R.}, {Cullen, F.}, {Davé, R.}, {Dayal, P.}, {Ferrara, A.}, {Finlator, K.}, {Fynbo, J. P. U.}, {Flury, S. R.}, {Gelli, V.}, {Gillman, S.}, {Gottumukkala, R.}, {Gould, K.}, {Greve, T. R.}, {Hardin, S. E.}, {Hsiao, T. Y.-Y}, {Hutter, A.}, {Jakobsson, P.}, {Killi, M.}, {Khosravaninezhad, N.}, {Laursen, P.}, {Lee, M. M.}, {Magdis, G. E.}, {Matthee, J.}, {Naidu, R. P.}, {Narayanan, D.}, {Pollock, C.}, {Prescott, M. K. M.}, {Rusakov, V.}, {Shuntov, M.}, {Sneppen, A.}, {Smit, R.}, {Tanvir, N. R.}, {Terp, C.}, {Toft, S.}, {Valentino, F.}, {Vijayan, A. P.}, {Weaver, J. R.}, {Wise, J. H.}, \& {Witstok, J.}}]{Heintz2025}
{Heintz, K. E.}, {Brammer, G. B.}, {Watson, D.}, {et~al.} 2025, A\&A, 693, A60, \dodoi{10.1051/0004-6361/202450243}

\bibitem[{Helton {et~al.}(2024)Helton, Rieke, Alberts, Wu, Eisenstein, Hainline, Carniani, Ji, Baker, Bhatawdekar, Bunker, Cargile, Charlot, Chevallard, D'Eugenio, Egami, Johnson, Jones, Lyu, Maiolino, Pérez-González, Rieke, Robertson, Saxena, Scholtz, Shivaei, Sun, Tacchella, Whitler, Williams, Willmer, Willott, Witstok, \& Zhu}]{helton2024}
Helton, J.~M., Rieke, G.~H., Alberts, S., {et~al.} 2024.
\newblock \doarXiv{2405.18462}

\bibitem[{Hunter {et~al.}(2023)Hunter, Indebetouw, Brogan, Berry, Chang, Francke, Geers, Gómez, Hibbard, Humphreys, Kent, Kepley, Kunneriath, Lipnicky, Loomis, Mason, Masters, Maud, Muders, Sabater, Sugimoto, Szűcs, Vasiliev, Videla, Villard, Williams, Xue, \& Yoon}]{Hunter_2023}
Hunter, T.~R., Indebetouw, R., Brogan, C.~L., {et~al.} 2023, Publications of the Astronomical Society of the Pacific, 135, 074501, \dodoi{10.1088/1538-3873/ace216}

\bibitem[{{Inami} {et~al.}(2022){Inami}, {Algera}, {Schouws}, {Sommovigo}, {Bouwens}, {Smit}, {Stefanon}, {Bowler}, {Endsley}, {Ferrara}, {Oesch}, {Stark}, {Aravena}, {Barrufet}, {da Cunha}, {Dayal}, {De Looze}, {Fudamoto}, {Gonzalez}, {Graziani}, {Hodge}, {Hygate}, {Nanayakkara}, {Pallottini}, {Riechers}, {Schneider}, {Topping}, \& {van der Werf}}]{inami2022}
{Inami}, H., {Algera}, H. S.~B., {Schouws}, S., {et~al.} 2022, \mnras, 515, 3126, \dodoi{10.1093/mnras/stac1779}

\bibitem[{{Isobe} {et~al.}(2023){Isobe}, {Ouchi}, {Nakajima}, {Harikane}, {Ono}, {Xu}, {Zhang}, \& {Umeda}}]{isobe2023}
{Isobe}, Y., {Ouchi}, M., {Nakajima}, K., {et~al.} 2023, \apj, 956, 139, \dodoi{10.3847/1538-4357/acf376}

\bibitem[{{Jones} {et~al.}(2020){Jones}, {Sanders}, {Roberts-Borsani}, {Ellis}, {Laporte}, {Treu}, \& {Harikane}}]{JonesTucker2020}
{Jones}, T., {Sanders}, R., {Roberts-Borsani}, G., {et~al.} 2020, \apj, 903, 150, \dodoi{10.3847/1538-4357/abb943}

\bibitem[{{Kaasinen} {et~al.}(2023){Kaasinen}, {van Marrewijk}, {Popping}, {Ginolfi}, {Di Mascolo}, {Mroczkowski}, {Concas}, {Di Cesare}, {Killi}, \& {Langan}}]{kaasinen2023}
{Kaasinen}, M., {van Marrewijk}, J., {Popping}, G., {et~al.} 2023, \aap, 671, A29, \dodoi{10.1051/0004-6361/202245093}

\bibitem[{{Katz} {et~al.}(2017){Katz}, {Kimm}, {Sijacki}, \& {Haehnelt}}]{katz2017}
{Katz}, H., {Kimm}, T., {Sijacki}, D., \& {Haehnelt}, M.~G. 2017, \mnras, 468, 4831, \dodoi{10.1093/mnras/stx608}

\bibitem[{Katz {et~al.}(2019)Katz, Galligan, Kimm, Rosdahl, Haehnelt, Blaizot, Devriendt, Slyz, Laporte, \& Ellis}]{Katz_2019}
Katz, H., Galligan, T.~P., Kimm, T., {et~al.} 2019, Monthly Notices of the Royal Astronomical Society, 487, 5902–5921, \dodoi{10.1093/mnras/stz1672}

\bibitem[{{Katz} {et~al.}(2022){Katz}, {Rosdahl}, {Kimm}, {Garel}, {Blaizot}, {Haehnelt}, {Michel-Dansac}, {Martin-Alvarez}, {Devriendt}, {Slyz}, {Teyssier}, {Ocvirk}, {Laporte}, \& {Ellis}}]{katz2022}
{Katz}, H., {Rosdahl}, J., {Kimm}, T., {et~al.} 2022, \mnras, 510, 5603, \dodoi{10.1093/mnras/stac028}

\bibitem[{{Kepley} {et~al.}(2020){Kepley}, {Tsutsumi}, {Brogan}, {Indebetouw}, {Yoon}, {Mason}, \& {Donovan Meyer}}]{kepley2020}
{Kepley}, A.~A., {Tsutsumi}, T., {Brogan}, C.~L., {et~al.} 2020, \pasp, 132, 024505, \dodoi{10.1088/1538-3873/ab5e14}

\bibitem[{{Kohandel} {et~al.}(2023){Kohandel}, {Ferrara}, {Pallottini}, {Vallini}, {Sommovigo}, \& {Ziparo}}]{kohandel2023}
{Kohandel}, M., {Ferrara}, A., {Pallottini}, A., {et~al.} 2023, \mnras, 520, L16, \dodoi{10.1093/mnrasl/slac166}

\bibitem[{{Le F{\`e}vre} {et~al.}(2020){Le F{\`e}vre}, {B{\'e}thermin}, {Faisst}, {Jones}, {Capak}, {Cassata}, {Silverman}, {Schaerer}, {Yan}, {Amorin}, {Bardelli}, {Boquien}, {Cimatti}, {Dessauges-Zavadsky}, {Giavalisco}, {Hathi}, {Fudamoto}, {Fujimoto}, {Ginolfi}, {Gruppioni}, {Hemmati}, {Ibar}, {Koekemoer}, {Khusanova}, {Lagache}, {Lemaux}, {Loiacono}, {Maiolino}, {Mancini}, {Narayanan}, {Morselli}, {M{\'e}ndez-Hern{\`a}ndez}, {Oesch}, {Pozzi}, {Romano}, {Riechers}, {Scoville}, {Talia}, {Tasca}, {Thomas}, {Toft}, {Vallini}, {Vergani}, {Walter}, {Zamorani}, \& {Zucca}}]{lefevre2020}
{Le F{\`e}vre}, O., {B{\'e}thermin}, M., {Faisst}, A., {et~al.} 2020, \aap, 643, A1, \dodoi{10.1051/0004-6361/201936965}

\bibitem[{{Liang} {et~al.}(2019){Liang}, {Feldmann}, {Kere{\v{s}}}, {Scoville}, {Hayward}, {Faucher-Gigu{\`e}re}, {Schreiber}, {Ma}, {Hopkins}, \& {Quataert}}]{liang2019}
{Liang}, L., {Feldmann}, R., {Kere{\v{s}}}, D., {et~al.} 2019, \mnras, 489, 1397, \dodoi{10.1093/mnras/stz2134}

\bibitem[{{Markov} {et~al.}(2024){Markov}, {Gallerani}, {Ferrara}, {Pallottini}, {Parlanti}, {Di Mascia}, {Sommovigo}, \& {Kohandel}}]{Markov_2024}
{Markov}, V., {Gallerani}, S., {Ferrara}, A., {et~al.} 2024, arXiv e-prints, arXiv:2402.05996, \dodoi{10.48550/arXiv.2402.05996}

\bibitem[{{Maseda} {et~al.}(2017){Maseda}, {Brinchmann}, {Franx}, {Bacon}, {Bouwens}, {Schmidt}, {Boogaard}, {Contini}, {Feltre}, {Inami}, {Kollatschny}, {Marino}, {Richard}, {Verhamme}, \& {Wisotzki}}]{Maseda2017}
{Maseda}, M.~V., {Brinchmann}, J., {Franx}, M., {et~al.} 2017, \aap, 608, A4, \dodoi{10.1051/0004-6361/201730985}

\bibitem[{Meyer {et~al.}(2024)Meyer, Oesch, Giovinazzo, Weibel, Brammer, Matthee, Naidu, Bouwens, Chisholm, Covelo-Paz, Fudamoto, Maseda, Nelson, Shivaei, Xiao, Herard-Demanche, Illingworth, Kerutt, Kramarenko, Labbe, Leonova, Magee, Matharu, Prieto Lyon, Reddy, Schaerer, Shapley, Stefanon, Wozniak, \& Wuyts}]{meyer2024}
Meyer, R.~A., Oesch, P.~A., Giovinazzo, E., {et~al.} 2024, Monthly Notices of the Royal Astronomical Society, 535, 1067, \dodoi{10.1093/mnras/stae2353}

\bibitem[{{Micha{\l}owski}(2015)}]{michalowski2015}
{Micha{\l}owski}, M.~J. 2015, \aap, 577, A80, \dodoi{10.1051/0004-6361/201525644}

\bibitem[{Milisavljevic {et~al.}(2024)Milisavljevic, Temim, Looze, Dickinson, Laming, Fesen, Raymond, Arendt, Vink, Posselt, Pavlov, Fox, Pinarski, Subrayan, Schmidt, Blair, Rest, Patnaude, Koo, Rho, Orlando, Janka, Andrews, Barlow, Burrows, Chevalier, Clayton, Fransson, Fryer, Gomez, Kirchschlager, Lee, Matsuura, Niculescu-Duvaz, Pierel, Plucinsky, Priestley, Ravi, Sartorio, Schmidt, Shahbandeh, Slane, Smith, Sravan, Weil, Wesson, \& Wheeler}]{Milisavljevic_2024}
Milisavljevic, D., Temim, T., Looze, I.~D., {et~al.} 2024, The Astrophysical Journal Letters, 965, L27, \dodoi{10.3847/2041-8213/ad324b}

\bibitem[{Moriwaki {et~al.}(2018)Moriwaki, Yoshida, Shimizu, Harikane, Matsuda, Matsuo, Hashimoto, Inoue, Tamura, \& Nagao}]{Moriwaki_2018}
Moriwaki, K., Yoshida, N., Shimizu, I., {et~al.} 2018, Monthly Notices of the Royal Astronomical Society: Letters, 481, L84–L88, \dodoi{10.1093/mnrasl/sly167}

\bibitem[{{Naidu} {et~al.}(2022){Naidu}, {Oesch}, {van Dokkum}, {Nelson}, {Suess}, {Brammer}, {Whitaker}, {Illingworth}, {Bouwens}, {Tacchella}, {Matthee}, {Allen}, {Bezanson}, {Conroy}, {Labbe}, {Leja}, {Leonova}, {Magee}, {Price}, {Setton}, {Strait}, {Stefanon}, {Toft}, {Weaver}, \& {Weibel}}]{naidu2022}
{Naidu}, R.~P., {Oesch}, P.~A., {van Dokkum}, P., {et~al.} 2022, \apjl, 940, L14, \dodoi{10.3847/2041-8213/ac9b22}

\bibitem[{{Nakazato} {et~al.}(2023){Nakazato}, {Yoshida}, \& {Ceverino}}]{nakazato2023}
{Nakazato}, Y., {Yoshida}, N., \& {Ceverino}, D. 2023, \apj, 953, 140, \dodoi{10.3847/1538-4357/ace25a}

\bibitem[{Oesch {et~al.}(2016)Oesch, Brammer, Dokkum, Illingworth, Bouwens, Labbé, Franx, Momcheva, Ashby, Fazio, Gonzalez, Holden, Magee, Skelton, Smit, Spitler, Trenti, \& Willner}]{Oesch_2016}
Oesch, P.~A., Brammer, G., Dokkum, P. G.~v., {et~al.} 2016, The Astrophysical Journal, 819, 129, \dodoi{10.3847/0004-637x/819/2/129}

\bibitem[{{Oesch} {et~al.}(2023){Oesch}, {Brammer}, {Naidu}, {Bouwens}, {Chisholm}, {Illingworth}, {Matthee}, {Nelson}, {Qin}, {Reddy}, {Shapley}, {Shivaei}, {van Dokkum}, {Weibel}, {Whitaker}, {Wuyts}, {Covelo-Paz}, {Endsley}, {Fudamoto}, {Giovinazzo}, {Herard-Demanche}, {Kerutt}, {Kramarenko}, {Labbe}, {Leonova}, {Lin}, {Magee}, {Marchesini}, {Maseda}, {Mason}, {Matharu}, {Meyer}, {Neufeld}, {Prieto Lyon}, {Schaerer}, {Sharma}, {Shuntov}, {Smit}, {Stefanon}, {Wyithe}, \& {Xiao}}]{Oesch_2023}
{Oesch}, P.~A., {Brammer}, G., {Naidu}, R.~P., {et~al.} 2023, \mnras, 525, 2864, \dodoi{10.1093/mnras/stad2411}

\bibitem[{{Oke} \& {Gunn}(1983)}]{oke_gunn_1983ApJ...266..713O}
{Oke}, J.~B., \& {Gunn}, J.~E. 1983, \apj, 266, 713, \dodoi{10.1086/160817}

\bibitem[{{Pallottini} {et~al.}(2022){Pallottini}, {Ferrara}, {Gallerani}, {Behrens}, {Kohandel}, {Carniani}, {Vallini}, {Salvadori}, {Gelli}, {Sommovigo}, {D'Odorico}, {Di Mascia}, \& {Pizzati}}]{pallottini2022}
{Pallottini}, A., {Ferrara}, A., {Gallerani}, S., {et~al.} 2022, \mnras, 513, 5621, \dodoi{10.1093/mnras/stac1281}

\bibitem[{Popping(2023)}]{Popping_2023}
Popping, G. 2023, Astronomy \& Astrophysics, 669, L8, \dodoi{10.1051/0004-6361/202244831}

\bibitem[{Ravindranath {et~al.}(2020)Ravindranath, Monroe, Jaskot, Ferguson, \& Tumlinson}]{Ravindranath_2020}
Ravindranath, S., Monroe, T., Jaskot, A., Ferguson, H.~C., \& Tumlinson, J. 2020, The Astrophysical Journal, 896, 170, \dodoi{10.3847/1538-4357/ab91a5}

\bibitem[{Robertson {et~al.}(2023)Robertson, Tacchella, Johnson, Hainline, Whitler, Eisenstein, Endsley, Rieke, Stark, Alberts, Dressler, Egami, Hausen, Rieke, Shivaei, Williams, Willmer, Arribas, Bonaventura, Bunker, Cameron, Carniani, Charlot, Chevallard, Curti, Curtis-Lake, D’Eugenio, Jakobsen, Looser, Lützgendorf, Maiolino, Maseda, Rawle, Rix, Smit, Übler, Willott, Witstok, Baum, Bhatawdekar, Boyett, Chen, de~Graaff, Florian, Helton, Hviding, Ji, Kumari, Lyu, Nelson, Sandles, Saxena, Suess, Sun, Topping, \& Wallace}]{Robertson_2023}
Robertson, B.~E., Tacchella, S., Johnson, B.~D., {et~al.} 2023, Nature Astronomy, 7, 611–621, \dodoi{10.1038/s41550-023-01921-1}

\bibitem[{Santini {et~al.}(2023)Santini, Fontana, Castellano, Leethochawalit, Trenti, Treu, Belfiori, Birrer, Bonchi, Merlin, Mason, Morishita, Nonino, Paris, Polenta, Rosati, Yang, Boyett, Bradac, Calabrò, Dressler, Glazebrook, Marchesini, Mascia, Nanayakkara, Pentericci, Roberts-Borsani, Scarlata, Vulcani, \& Wang}]{Santini_2023}
Santini, P., Fontana, A., Castellano, M., {et~al.} 2023, The Astrophysical Journal Letters, 942, L27, \dodoi{10.3847/2041-8213/ac9586}

\bibitem[{{Schneider} \& {Maiolino}(2024)}]{schneider2024}
{Schneider}, R., \& {Maiolino}, R. 2024, \aapr, 32, 2, \dodoi{10.1007/s00159-024-00151-2}

\bibitem[{{Schouws} {et~al.}(2022){Schouws}, {Stefanon}, {Bouwens}, {Smit}, {Hodge}, {Labb{\'e}}, {Algera}, {Boogaard}, {Carniani}, {Fudamoto}, {Holwerda}, {Illingworth}, {Maiolino}, {Maseda}, {Oesch}, \& {van der Werf}}]{schouws2022}
{Schouws}, S., {Stefanon}, M., {Bouwens}, R., {et~al.} 2022, \apj, 928, 31, \dodoi{10.3847/1538-4357/ac4605}

\bibitem[{{Schouws} {et~al.}(2023){Schouws}, {Bouwens}, {Smit}, {Hodge}, {Stefanon}, {Witstok}, {Hilhorst}, {Labb{\'e}}, {Algera}, {Boogaard}, {Maseda}, {Oesch}, {R{\"o}ttgering}, \& {van der Werf}}]{Schouws_2022}
{Schouws}, S., {Bouwens}, R., {Smit}, R., {et~al.} 2023, \apj, 954, 103, \dodoi{10.3847/1538-4357/ace10c}

\bibitem[{{Schreiber} {et~al.}(2018){Schreiber}, {Elbaz}, {Pannella}, {Ciesla}, {Wang}, \& {Franco}}]{schreiber2018}
{Schreiber}, C., {Elbaz}, D., {Pannella}, M., {et~al.} 2018, \aap, 609, A30, \dodoi{10.1051/0004-6361/201731506}

\bibitem[{{Solomon} {et~al.}(1992){Solomon}, {Downes}, \& {Radford}}]{Solomon_1992}
{Solomon}, P.~M., {Downes}, D., \& {Radford}, S.~J.~E. 1992, \apjl, 387, L55, \dodoi{10.1086/186304}

\bibitem[{{Sommovigo} {et~al.}(2021){Sommovigo}, {Ferrara}, {Carniani}, {Zanella}, {Pallottini}, {Gallerani}, \& {Vallini}}]{sommovigo2021}
{Sommovigo}, L., {Ferrara}, A., {Carniani}, S., {et~al.} 2021, \mnras, 503, 4878, \dodoi{10.1093/mnras/stab720}

\bibitem[{{Sommovigo} {et~al.}(2022{\natexlab{a}}){Sommovigo}, {Ferrara}, {Pallottini}, {Dayal}, {Bouwens}, {Smit}, {da Cunha}, {De Looze}, {Bowler}, {Hodge}, {Inami}, {Oesch}, {Endsley}, {Gonzalez}, {Schouws}, {Stark}, {Stefanon}, {Aravena}, {Graziani}, {Riechers}, {Schneider}, {van der Werf}, {Algera}, {Barrufet}, {Fudamoto}, {Hygate}, {Labb{\'e}}, {Li}, {Nanayakkara}, \& {Topping}}]{sommovigo2022}
{Sommovigo}, L., {Ferrara}, A., {Pallottini}, A., {et~al.} 2022{\natexlab{a}}, \mnras, 513, 3122, \dodoi{10.1093/mnras/stac302}

\bibitem[{{Sommovigo} {et~al.}(2022{\natexlab{b}}){Sommovigo}, {Ferrara}, {Carniani}, {Pallottini}, {Dayal}, {Pizzati}, {Ginolfi}, {Markov}, \& {Faisst}}]{sommovigo2022b}
{Sommovigo}, L., {Ferrara}, A., {Carniani}, S., {et~al.} 2022{\natexlab{b}}, \mnras, 517, 5930, \dodoi{10.1093/mnras/stac2997}

\bibitem[{{Tang} {et~al.}(2021){Tang}, {Stark}, {Chevallard}, {Charlot}, {Endsley}, \& {Congiu}}]{Tang2021}
{Tang}, M., {Stark}, D.~P., {Chevallard}, J., {et~al.} 2021, \mnras, 501, 3238, \dodoi{10.1093/mnras/staa3454}

\bibitem[{{Thompson} {et~al.}(2017){Thompson}, {Moran}, \& {Swenson}}]{thompson2017}
{Thompson}, A.~R., {Moran}, J.~M., \& {Swenson}, George~W., J. 2017, {Interferometry and Synthesis in Radio Astronomy, 3rd Edition}, \dodoi{10.1007/978-3-319-44431-4}

\bibitem[{Treu {et~al.}(2023)Treu, Calabrò, Castellano, Leethochawalit, Merlin, Fontana, Yang, Morishita, Trenti, Dressler, Mason, Paris, Pentericci, Roberts-Borsani, Vulcani, Boyett, Bradac, Glazebrook, Jones, Marchesini, Mascia, Nanayakkara, Santini, Strait, Vanzella, \& Wang}]{Treu_2023}
Treu, T., Calabrò, A., Castellano, M., {et~al.} 2023, The Astrophysical Journal Letters, 942, L28, \dodoi{10.3847/2041-8213/ac9283}

\bibitem[{{Vallini} {et~al.}(2021){Vallini}, {Ferrara}, {Pallottini}, {Carniani}, \& {Gallerani}}]{vallini2021}
{Vallini}, L., {Ferrara}, A., {Pallottini}, A., {Carniani}, S., \& {Gallerani}, S. 2021, \mnras, 505, 5543, \dodoi{10.1093/mnras/stab1674}

\bibitem[{{Vallini} {et~al.}(2024){Vallini}, {Witstok}, {Sommovigo}, {Pallottini}, {Ferrara}, {Carniani}, {Kohandel}, {Smit}, {Gallerani}, \& {Gruppioni}}]{vallini2024}
{Vallini}, L., {Witstok}, J., {Sommovigo}, L., {et~al.} 2024, \mnras, 527, 10, \dodoi{10.1093/mnras/stad3150}

\bibitem[{{Weingartner} \& {Draine}(2001)}]{weingartner2001}
{Weingartner}, J.~C., \& {Draine}, B.~T. 2001, \apj, 548, 296, \dodoi{10.1086/318651}

\bibitem[{{Williams} {et~al.}(2024){Williams}, {Alberts}, {Ji}, {Hainline}, {Lyu}, {Rieke}, {Endsley}, {Suess}, {Sun}, {Johnson}, {Florian}, {Shivaei}, {Rujopakarn}, {Baker}, {Bhatawdekar}, {Boyett}, {Bunker}, {Cameron}, {Carniani}, {Charlot}, {Curtis-Lake}, {DeCoursey}, {de Graaff}, {Egami}, {Eisenstein}, {Gibson}, {Hausen}, {Helton}, {Maiolino}, {Maseda}, {Nelson}, {P{\'e}rez-Gonz{\'a}lez}, {Rieke}, {Robertson}, {Saxena}, {Tacchella}, {Willmer}, \& {Willott}}]{Williams_2023}
{Williams}, C.~C., {Alberts}, S., {Ji}, Z., {et~al.} 2024, \apj, 968, 34, \dodoi{10.3847/1538-4357/ad3f17}

\bibitem[{{Witstok} {et~al.}(2023{\natexlab{a}}){Witstok}, {Jones}, {Maiolino}, {Smit}, \& {Schneider}}]{Witstok23_dust}
{Witstok}, J., {Jones}, G.~C., {Maiolino}, R., {Smit}, R., \& {Schneider}, R. 2023{\natexlab{a}}, \mnras, 523, 3119, \dodoi{10.1093/mnras/stad1470}

\bibitem[{{Witstok} {et~al.}(2022){Witstok}, {Smit}, {Maiolino}, {Kumari}, {Aravena}, {Boogaard}, {Bouwens}, {Carniani}, {Hodge}, {Jones}, {Stefanon}, {van der Werf}, \& {Schouws}}]{Witstok_2022}
{Witstok}, J., {Smit}, R., {Maiolino}, R., {et~al.} 2022, \mnras, 515, 1751, \dodoi{10.1093/mnras/stac1905}

\bibitem[{{Witstok} {et~al.}(2023{\natexlab{b}}){Witstok}, {Shivaei}, {Smit}, {Maiolino}, {Carniani}, {Curtis-Lake}, {Ferruit}, {Arribas}, {Bunker}, {Cameron}, {Charlot}, {Chevallard}, {Curti}, {de Graaff}, {D'Eugenio}, {Giardino}, {Looser}, {Rawle}, {Rodr{\'\i}guez del Pino}, {Willott}, {Alberts}, {Baker}, {Boyett}, {Egami}, {Eisenstein}, {Endsley}, {Hainline}, {Ji}, {Johnson}, {Kumari}, {Lyu}, {Nelson}, {Perna}, {Rieke}, {Robertson}, {Sandles}, {Saxena}, {Scholtz}, {Sun}, {Tacchella}, {Williams}, \& {Willmer}}]{Witstok2023_JWST}
{Witstok}, J., {Shivaei}, I., {Smit}, R., {et~al.} 2023{\natexlab{b}}, \nat, 621, 267, \dodoi{10.1038/s41586-023-06413-w}

\bibitem[{Yang \& Lidz(2020)}]{Yang_2020}
Yang, S., \& Lidz, A. 2020, Monthly Notices of the Royal Astronomical Society, 499, 3417–3433, \dodoi{10.1093/mnras/staa3000}

\bibitem[{Yoon {et~al.}(2023)Yoon, Carilli, Fujimoto, Castellano, Merlin, Santini, Yun, Murphy, Jung, Casey, Finkelstein, Papovich, Fontana, Treu, \& Letai}]{Yoon_2023}
Yoon, I., Carilli, C.~L., Fujimoto, S., {et~al.} 2023, The Astrophysical Journal, 950, 61, \dodoi{10.3847/1538-4357/acc94d}

\bibitem[{{Zavala} {et~al.}(2024){Zavala}, {Castellano}, {Akins}, {Bakx}, {Burgarella}, {Casey}, {Ch{\'a}vez Ortiz}, {Dickinson}, {Finkelstein}, {Mitsuhashi}, {Nakajima}, {P{\'e}rez-Gonz{\'a}lez}, {Arrabal Haro}, {Buat}, {Backhaus}, {Calabr{\`o}}, {Cleri}, {Fern{\'a}ndez-Arenas}, {Fontana}, {Franco}, {Giavalisco}, {Grogin}, {Hathi}, {Hirschmann}, {Ikeda}, {Jung}, {Kartaltepe}, {Koekemoer}, {Larson}, {McKinney}, {Papovich}, {Saito}, {Santini}, {Terlevich}, {Terlevich}, {Treu}, \& {Yung}}]{zavala2024}
{Zavala}, J.~A., {Castellano}, M., {Akins}, H.~B., {et~al.} 2024, arXiv e-prints, arXiv:2403.10491, \dodoi{10.48550/arXiv.2403.10491}

\end{thebibliography}
\bibliographystyle{aasjournal}

\end{document}